# Forecasting CPI Inflation Components with Hierarchical Recurrent Neural Networks

Oren Barkan[a], Jonathan Benchimol[b], Itamar Caspi[b], Eliya Cohen[c], Allon Hammer[c], Noam Koenigstein[c,1]

[a] *Ariel University*
[b] *Bank of Israel*
[c] *Tel-Aviv University*

---

**Abstract**

We present a hierarchical architecture based on recurrent neural networks for predicting disaggregated inflation components of the Consumer Price Index (CPI). While the majority of existing research is focused on predicting headline inflation, many economic and financial institutions are interested in its partial disaggregated components. To this end, we developed the novel Hierarchical Recurrent Neural Network (HRNN) model, which utilizes information from higher levels in the CPI hierarchy to improve predictions at the more volatile lower levels. Based on a large dataset from the US CPI-U index, our evaluations indicate that the HRNN model significantly outperforms a vast array of well-known inflation prediction baselines. Our methodology and results provide additional forecasting measures and possibilities to policy and market makers on sectoral and component-specific price changes.

*Keywords:* Inflation forecasting, Disaggregated inflation, Consumer Price Index, Machine learning, Gated Recurrent Unit, Recurrent Neural Networks.

*JEL Classification:* C45, C53, E31, E37.

---

*Email address:* `noamk@tauex.tau.ac.il` (Noam Koenigstein)

## 1. Introduction

The Consumer Price Index (CPI) is a measure of the average change over time in the prices paid by a representative consumer for a common basket of goods and services. The CPI attempts to quantify and measure the average cost-of-living in a given country by estimating the purchasing power of a single unit of currency. Therefore, it is the key macroeconomic indicator for measuring inflation (or deflation). As such, the CPI is a major driving force in the economy influencing a plethora of market dynamics. In this work, we present a novel model based on Recurrent Neural Networks (RNNs) for forecasting disaggregated CPI inflation components.

In the mid-1980s, many advanced economies began a major process of disinflation known as the "great moderation". This period was characterized by steady low inflation and moderate yet steady economic growth (Faust and Wright, 2013). Later, the Global Financial Crisis (GFC) of 2008, and more recently the economic effects of the Covid-19 pandemic, were met with unprecedented monetary policies, potentially altering the underlying inflation dynamics worldwide (Woodford, 2012; Gilchrist et al., 2017; Bernanke et al., 2018). While economists still debate about the underlying forces that drive inflation, all agree on the importance and value of contemporary inflation research, measurements and estimation. Moreover, the CPI is a composite index comprised of an elaborate hierarchy of sub-indexes each with its own dynamics and driving forces. Hence, in order to better understand inflation dynamics, it is useful to deconstruct the CPI index and look into the specific disaggregated components "underneath" the main headline.

In the US, the Consumer Price Index (CPI) is calculated and reported by the Bureau of Labor Statistics (BLS). It represents the cost of a basket of goods and services across the country on a monthly basis. The CPI is a hierarchical composite index system that partitions all consumer goods and services into a hierarchy of increasingly detailed categories. In the US, the top CPI headline is composed of eight major sector indexes: (1) Housing, (2) Food and Beverages, (3) Medical Care, (4) Apparel, (5) Transportation, (6) Energy, (7) Recreation, and (8) Other goods and services. Each sector is composed of finer and finer sub-indexes until the entry-levels or "leaves" are reached. These entry -level indexes represent concrete measurable products or services whose price levels are being tracked. For example, the *White Bread* entry is classified under the following eight-level hierarchy: *All Items* → *Food and Beverages* → *Food at Home* → *Cereals and Bakery Products* → *Cereals and Cereal Products* → *Bakery products* → *Bread* → *White Bread*.

The ability to accurately estimate the upcoming disaggregated inflation rate is of high interest to policymakers and market players: Inflation forecasting is a critical tool in adjusting monetary policies around the world (Friedman, 1961). Central banks predict future inflation trends to justify interest rate decisions and to control and maintain inflation around its target. Better understanding of upcoming inflation dynamics at the component level can help inform and elucidate decision-makers for optimal monetary policy (Ida, 2020). Predicting disaggregated inflation rates is also important to fiscal authorities that wish to forecast sectoral inflation dynamics to adjust social security payments and assistance packages to specific industrial sectors. In the private sector, investors in fixed-income markets wish to estimate future sectorial inflation in order to foresee upcoming trends in discounted real returns. Additionally, some private firms need to predict specific inflation components in order to forecast price dynamics and mitigate risks accordingly. Finally, both government and private debt levels and interest payments heavily depend on the expected path of inflation. These are just a few examples that emphasize the importance of disaggregated inflation forecasting.

Most existing inflation forecasting models attempt to predict the headline CPI while implicitly assuming the same approach can be effectively applied to its disaggregated components (Faust and Wright, 2013). However, as we show later, and in line with the literature, the disaggregated components are more volatile and harder to predict. Moreover, changes in the CPI components are more prevalent at the lower levels than up at the main categories. As a result, lower hierarchy levels often have less historical measurements for training modern machine learning algorithms.

In this work, we present the Hierarchical Recurrent Neural Network (HRNN) model, a novel model based on RNNs that utilizes the CPI's inherent hierarchy for improved predictions at its lower levels. HRNN is a hierarchical arrangement of RNNs analogous to the CPI's hierarchy. This architecture allows information to propagate from higher to lower levels in order to mitigate volatility and information sparsity that otherwise impedes advanced machine learning approaches. Hence, a key advantage of the HRNN model stems from its superiority at inflation predictions at lower levels of the CPI hierarchy. Our evaluations indicate that HRNN outperforms many existing baselines at inflation forecasting of different CPI components below the top headline and across different time horizons.

Finally, our data and code are publicly available on GitHub[1] to enable reproducibility and foster future evaluations of new methods. By doing so, we comply with the call to make data and algorithms more open and transparent to the community (Makridakis et al., 2018, 2020).

The remainder of the paper is organized as follows. Section 2 presents a literature review of baseline inflation forecasting models and machine learning models. Section 3 explains recurrent neural networks methodologies. Our novel HRNN model is presented in Section 4. Section 5 describes the price data and data transformations. In Section 6, we present our results and compare them to alternative approaches. Finally, we conclude in Section 7 by discussing potential implications of the current research and several future directions.

## 2. Related Work

While inflation forecasting is a challenging task of high importance, the literature indicates that significant improvement upon basic time-series models and heuristics is hard to achieve. Indeed, Atkeson et al. (2001) found that forecasts based on simple averages of past inflation were more accurate than all other alternatives, including the canonical Phillips curve and other forms of structural models. Similarly, Stock and Watson (2007, 2010) provide empirical evidence for the superiority of univariate models in forecasting inflation during the great moderation period (1985 to 2007) and during the recovery ensuing the GFC. More recently, Faust and Wright (2013) conducted an extensive survey of inflation forecasting methods and found that a simple "glide path" prediction from the current inflation rate performs as well as model-based forecasts for long-run inflation rates and often outperforms them.

[1] The code and data are available at `https://github.com/AllonHammer/CPI_HRNN`

Recently, an increasing amount of effort has been directed towards the application of machine learning models for inflation forecasting. For example, Medeiros et al. (2021) compared inflation forecasting with several machine learning models such as lasso regression, random forests, and deep neural networks. However, Medeiros et al. (2021) mainly focused on using exogenous features such as cash and credit availability, online prices, housing prices, consumer data, exchange rates, and interest rates. When exogenous features are considered, the emphasis shifts from learning the endogenous time series patterns to effectively extracting the predictive information from the exogenous features. In contrast to Medeiros et al. (2021), we preclude the use of any exogenous features and focus on harnessing the internal patterns of the CPI series. Moreover, unlike previous works that dealt with estimating the main headline, this work is focused on predicting the disaggregated indexes that comprise the CPI.

In general, machine learning methods flourish where data is found in abundance and many *training* examples are available. Unfortunately, this is not the case with CPI inflation data. While a large amount of relevant exogenous features exist, there are only twelve monthly readings annually. Hence, the amount of available training examples is limited. Furthermore, Stock and Watson (2007) show that statistics such as average inflation rate, conditional volatility, and persistency levels are shifting in time. Hence, inflation is a *non-stationary* process, which further limits the amount of relevant historical data points.

Goulet Coulombe et al. (2022), Mullainathan and Spiess (2017), Athey and Susan (2018) and Chakraborty and Joseph (2017) present comprehensive surveys of general machine learning applications in economics. Here, we do not attempt to cover the plethora of research employing machine learning for economic forecasting. Instead, we focus on models that apply neural networks to CPI forecasting in the next section.

This paper joins several studies that apply neural network methods to the specific task of inflation forecasting: Nakamura (2005) employed a simple feed-forward network to predict quarterly CPI headline values. A special emphasis is placed on *early stopping* methodologies in order to prevent over-fitting. Their evaluations are based on US CPI data during 1978-2003 and predictions are compared against several autoregressive (AR) baselines. Presented in Section 6, our evaluations confirm the findings of Nakamura (2005), that a fully connected network is indeed effective at predicting the headline CPI. However, when the CPI components are considered, we show that the model in this work demonstrates superior accuracy.

Choudhary and Haider (2012) used several neural networks to forecast monthly inflation rates in 28 countries in the Organisation for Economic Cooperation and Development (OECD). Their findings showed that, on average, neural network models were superior in 45% of the countries while simple AR models of order one (AR1) performed better in 23% of the countries. They also proposed to combine an ensemble of multiple networks arithmetically for further accuracy.

Chen et al. (2001) explored semi-parametric nonlinear autoregressive models with exogenous variables (NLARX) based on neural networks. Their investigation covered a comparison of different nonlinear activation functions such as the Sigmoid activation, radial basis activation, and Ridgelet activation.

McAdam and McNelis (2005) explored Thick Neural Network models that represent "trimmed mean" forecasts from several models. By combining the network with a linear Phillips Curve model, they predict the CPI for the US, Japan, and Europe at different levels.

In contrast to the aforementioned works, our model predicts monthly CPI values in *all* hierarchy levels. We utilize information patterns from higher levels of the CPI hierarchy in order to assist the predictions at lower levels. Such predictions are more challenging due to the inherent noise and information sparsity at the lower levels. Moreover, the HRNN model in this work is better equipped to harness sequential patterns in the data by employing *Recurrent Neural Networks*. Finally, we exclude the use of exogenous variables and rely solely on historical CPI data to focus on internal CPI patterns modeling.

Almosova and Andresen (2019) employed long-short term memory LSTMs for inflation forecasting. They compared their approach to multiple baselines such as autoregressive models, random walk models, seasonal autoregressive models, Markov switching models, and fully-connected neural networks. At all time horizons, the root mean squared forecast of their LSTM model was approximately one-third of the random walk model and significantly more accurate than the other baselines.

As we explain in Section 3.3, our model uses Gated Recurrent Networks (GRUs), which are similar to LSTMs. Unlike Almosova and Andresen (2019) and Zahara et al. (2020), a key contribution of our model stems from its ability to propagate useful information from higher levels in the hierarchy down to the nodes at lower levels. By ignoring the hierarchical relations between the different CPI components, our model is reduced to a set of simple unrelated GRUs. This setup is similar to Almosova and Andresen (2019), as the difference between LSTMs and GRUs is negligible. In Section 6, we perform an ablation study in which HRNN ignores the hierarchical relations and is reduced to a collection of independent GRUs, similar to the model in Almosova and Andresen (2019). Our evaluations indicate that this approach is not optimal at any level of the CPI hierarchy.

## 3. Recurrent Neural Networks

Before describing the HRNN model in detail, we briefly overview the main different RNNs approaches. RNNs are neural networks that model sequences of data in which each value is assumed to be dependent on previous values. Specifically, RNNs are feed-forward networks augmented by implementing a feedback loop (Mandic and Chambers, 2001). As such, RNNs introduce a notion of time to the standard feed-forward neural networks and excel at modeling temporal dynamic behavior (Chung et al., 2014). Some RNN units retain an internal memory state from previous time steps representing an arbitrarily long context window. Many RNN implementations were proposed and studied in the past. A comprehensive review and comparison of the different RNN architectures is available in (Lipton et al., 2015) and (Chung et al., 2014). In this section, we will cover the three most popular units: Basic RNN, Long-Short Time Memory (LSTM), and Gated Recurrent Unit (GRU).

**Figure 1.** *An illustration of a basic RNN unit.*

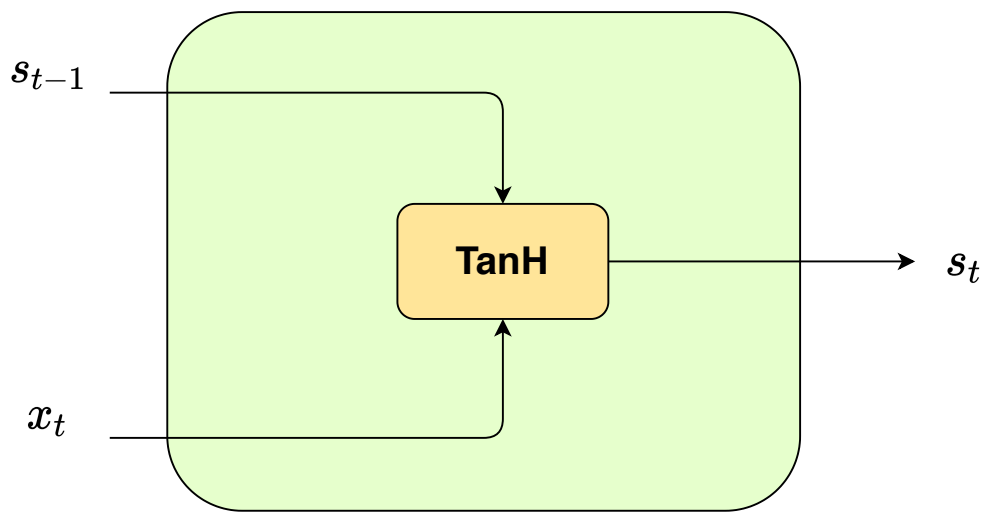

*Each line carries an entire vector, from the output of one node to the inputs of others. The yellow box is a learned neural network layer.*

### 3.1. *Basic Recurrent Neural Networks*

Let $\{x_t\}_{t=1}^{T}$ be the model's input time series consisting of $T$ samples. Similarly, let $\{s_t\}_{t=1}^{T}$ be the model's results consisting of $T$ samples from the target time series. Namely, the model's input at $t$ is $x_t$, and its output (prediction) is $s_t$. The following set of equations defines a basic RNN unit:

$$s_t = \tanh\left(x_t u + s_{t-1} w + b\right), \tag{1}$$

where $u$, $w$ and $b$ are the model's parameters and $\tanh(x) = \frac{e^x - e^{-x}}{e^x + e^{-x}}$ is the hyperbolic tangent function. Namely, the model's output from the previous period $s_{t-1}$ is used as an additional input to the model at time $t$, along with the current input $x_t$. The linear combination $x_t u + s_{t-1} w + b$ is the argument of a hyperbolic tangent *activation* function allowing the unit to model nonlinear relations between inputs and outputs. Different implementations may employ other activation functions, e.g., the Sigmoid function, some logistic functions, or a Rectified Linear Unit (ReLU) function (Ramachandran et al., 2017). Figure 1 depicts an illustration of a basic RNN unit.

### 3.2. *Long Short Term Memory Networks*

Basic RNNs suffer from the "short-term memory" problem: they utilize data from recent history to forecast, but if a sequence is long enough, it cannot carry relevant information from earlier periods to later ones, e.g., relevant patterns from the same month in previous years. Long Short Term Memory networks (LSTMs) mitigate the "short-term memory" problem by introducing gates that enable the preservation of relevant "long-term memory" and combining it with the most recent data (Hochreiter and Schmidhuber, 1997). The introduction of LSTMs paved the way for significant strides forward in various fields such as natural language processing, speech recognition, robot control, and more (Yu et al., 2019).

An LSTM unit has the ability to "memorize" or "forget" information through the use of a special memory *cell state*, carefully regulated by three gates: an *input gate*, a *forget gate*, and an *output gate*. The gates regulate the flow of information into and out of the memory cell state. An LSTM unit is defined by the following set of equations:

$$\begin{aligned} i &= \sigma(x_t u^i + s_{t-1} w^i + b^i), \\ f &= \sigma(x_t u^f + s_{t-1} w^f + b^f), \\ o &= \sigma(x_t u^o + s_{t-1} w^o + b^o), \\ \tilde{c} &= \tanh(x_t u^c + s_{t-1} w^c + b^c), \\ c_t &= f \times c_{t-1} + i \times \tilde{c}, \\ s_t &= o \times \tanh(c_t), \end{aligned} \tag{2}$$

where $\sigma(x) = \frac{1}{1+e^{-x}}$ is the sigmoid or logistic activation function. $u^i$, $w^i$ and $b^i$ are the learned parameters that control the *input gate* $i$. $u^f$, $w^f$ and $b^f$ are the learned parameters that control the *forget gate* $f$, and $u^o$, $w^o$ and $b^o$ are the learned parameters that control the *output gate* $o$. $\tilde{c}$ is the new candidate activation for the cell state determined by the parameters $u^c$, $w^c$ and $b^c$. The *cell state* itself $c_t$ is updated by the linear combination $c_t = f \times c_{t-1} + i \times \tilde{c}$, where $c_{t-1}$ is its previous value of the cell state. The input gate $i$ determines which parts of the candidate $\tilde{c}$ should be used to modify the memory cell state, and the forget gate $f$ determines which parts of the previous memory $c_{t-1}$ should be discarded. Finally, the recently updated *cell state* $c_t$ is "squashed" through a nonlinear hyperbolic tangent and the *output gate* $o$ determines which parts of it should be presented in the output $s_t$. Figure 2 depicts an illustration of an LSTM unit.

**Figure 2.** *An illustration of an LSTM Unit.*

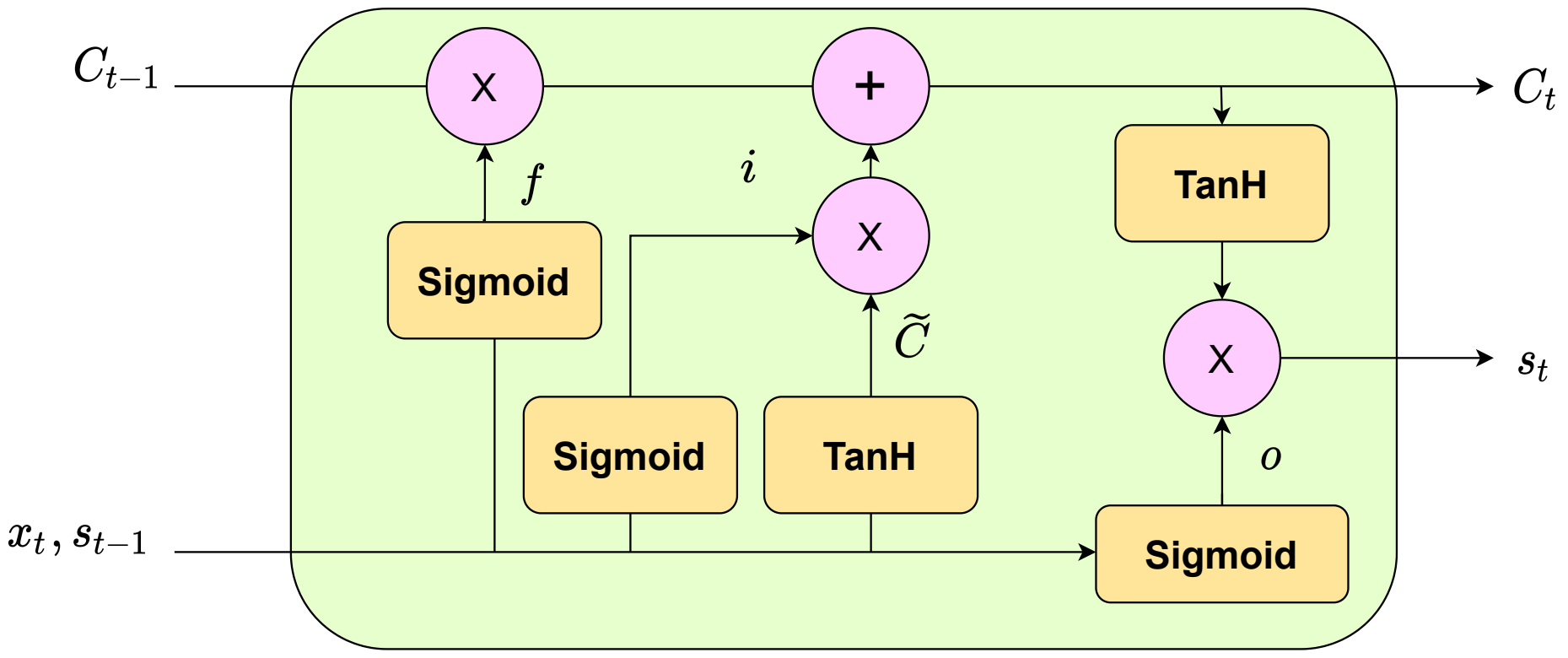

*Each line carries an entire vector, from the output of one node to the inputs of others. The pink circles represent point-wise operations, while the yellow boxes are learned neural network layers. Lines merging denote concatenation, while a line forking denotes its content being copied and the copies going to different locations.*

### 3.3. *Gated Recurrent Unit*

A Gated Recurrent Unit (GRU) improves the LSTM unit by dropping the *cell state* in favor of a more simplified unit that requires less learnable parameters (Dey and Salemt, 2017). GRU employs only two gates instead of three: an *update gate* and a *reset gate*. Using fewer parameters, GRUs are faster and more efficient, especially when training data is limited, such as in the case of inflation predictions and particularly disaggregated inflation components.

The following set of equations defines a GRU unit:

$$\begin{aligned} z &= \sigma(x_t u^z + s_{t-1} w^z + b^z), \\ r &= \sigma(x_t u^r + s_{t-1} w^r + b^r), \\ v &= \tanh\left(x_t u^v + (s_{t-1} \times r) w^v + b^v\right), \\ s_t &= z \times v + (1 - z) s_{t-1}, \end{aligned} \tag{3}$$

where $u^z$, $w^z$ and $b^z$ are the learned parameters that control the *update gate* $z$, and $u^r$, $w^r$ and $b^r$ are the learned parameters that control the *reset gate* $r$. The candidate activation $v$ is a function of the input $x_t$ and the previous output $s_{t-1}$, and is controlled by the learned parameters: $u^v$, $w^v$ and $b^v$. Finally, the output $s_t$ combines the candidate activation $v$ and the previous state $s_{t-1}$ controlled by the *update gate* $z$. Figure 2 depicts an illustration of a GRU unit.

GRUs enable the "memorization" of relevant information patterns with significantly fewer parameters compared to LSTMs. Hence, GRUs constitute the basic unit for our novel HRNN model described in Section 4.

**Figure 3.** *An illustration of a GRU unit.*

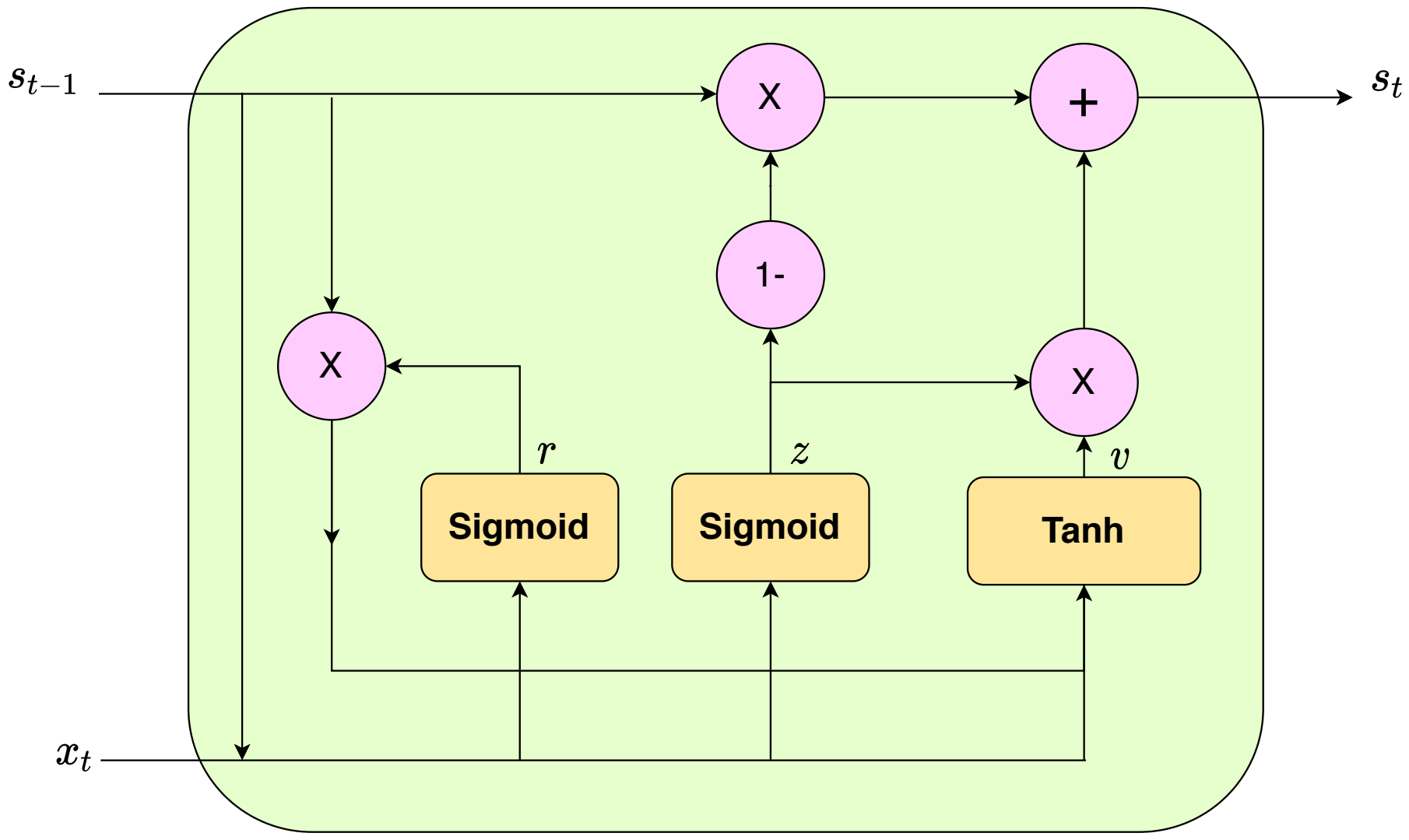

*Each line carries an entire vector, from the output of one node to the inputs of others. The pink circles represent point-wise operations, while the yellow boxes are learned neural network layers. Lines merging denote concatenation, while a line forking denotes its content being copied and the copies going to different locations.*

## 4. Hierarchical Recurrent Neural Networks

The disaggregated components at lower levels of the CPI hierarchy (e.g., newspapers, medical care, etc.) suffer from missing data as well as higher volatility in change rates. HRNN exhibits a network graph in which each node is associated with a RNN unit that models the inflation rate of a specific (sub)-index (node) in the "full" CPI hierarchy.

HRNN's unique architecture allows it to propagate information from RNN nodes in higher levels to lower levels in the CPI hierarchy, coarse to fine-grained, via a chain of hierarchical informative priors over the RNNs' parameters. This unique property of HRNN is materialized in better predictions for nodes at lower levels of the hierarchy, as we show later in Section 6,

### *4.1. Model Formulation*

Let $\mathcal{I} = \{n\}_{n=1}^{N}$ be an enumeration of the nodes in the CPI hierarchy graph. In addition, we define $\pi_n \in \mathcal{I}$ as the parent node of the node $n$. For example, if the nodes $n = 5$ and $n = 19$ represent the indexes of *tomatoes* and *vegetables* respectively, then $\pi_5 = 19$ i.e. the parent node of *tomatoes* is *vegetables*.

For each node $n \in \mathcal{I}$, we denote by $x_t^n \in \mathbb{R}$ the observed random variable that represents the CPI value of the node $n$ at timestamp $t \in \mathbb{N}$. We further denote $X_t^n \triangleq (x_1^n, ..., x_t^n)$, where $1 \leq t \leq T_n$ and $T_n$ is the last timestamp for node $n$. Let $g : \mathbb{R}^m \times \Omega \to \mathbb{R}$ be a parametric function representing an RNN node in the hierarchy. Specifically, $\mathbb{R}^m$ is the space of parameters that control the RNN unit, $\Omega$ is the input time series space, and the function $g$ predicts a scalar value for the next value of the input series. Hence, our goal is to learn the parameters $\theta_n \in \mathbb{R}^m$ s.t. for $X_t^n \in \Omega$, $g(\theta_n, X_t^n) = x_{t+1}^n$, $\forall n \in \mathcal{I}$, and $1 \leq t < T_n$.

We proceed by assuming a Gaussian error on $g$'s predictions and receive the following expression for the likelihood of the observed time series:

$$p(X_{T_n}^n | \theta_n, \tau_n) = \prod_{t=1}^{T_n} p(x_t^n | X_{t-1}^n, \theta_n, \tau_n) = \prod_{t=1}^{T_n} \mathcal{N}(x_t^n; g(\theta_n, X_{t-1}^n), \tau_n^{-1}), \tag{4}$$

where $\tau_n^{-1} \in \mathbb{R}$ is the variance of $g$'s errors.

Next, we define a hierarchical network of normal priors over the nodes' parameters that attach each node's parameters with those of its parent node. The hierarchical priors follow:

$$p(\theta_n | \theta_{\pi_n}, \tau_{\theta_n}) = \mathcal{N}(\theta_n; \theta_{\pi_n}, \tau_{\theta_n}^{-1}\mathbf{I}), \tag{5}$$

where $\tau_{\theta_n}$ is a configurable precision parameter that determines the "strength" of the relation between node $n$'s parameters and the parameters of its parent $\pi_n$. Higher values of $\tau_{\theta_n}$ strengthen the attachment between $\theta_n$ and its prior $\theta_{\pi_n}$.

The precision parameter $\tau_{\theta_n}$ can be seen as a global hyper-parameter of the model to be optimized via cross-validation. However, different nodes in CPI the hierarchy have varying degrees of correlation with their parent nodes. Hence, the value of $\tau_{\theta_n}$ in HRNN is given by:

$$\tau_{\theta_n} = e^{\alpha + C_n}, \tag{6}$$

where $\alpha$ is a hyper-parameter and $C_n = \rho(X_{T_n}^n, X_{T_{\pi_n}}^{\pi_n})$ is the Pearson correlation coefficient between the time series of $n$ and its parent $\pi_n$.

Importantly, Equation (5) describes a novel prior relationship between the parameters of a node and its parent in the hierarchy that "grows" increasingly stronger according to the historical correlation between the two series. This ensures that a child node $n$ is kept close to its parent node $\pi_n$ in terms of squared Euclidean distance in the parameters space, especially if they are highly correlated. Note that in the case of the root node (the headline CPI), $\pi_n$ does not exist and hence we set a normal non-informative regularization prior with zero mean and unit variance.

**Figure 4.** *An illustration of the full HRNN model.*

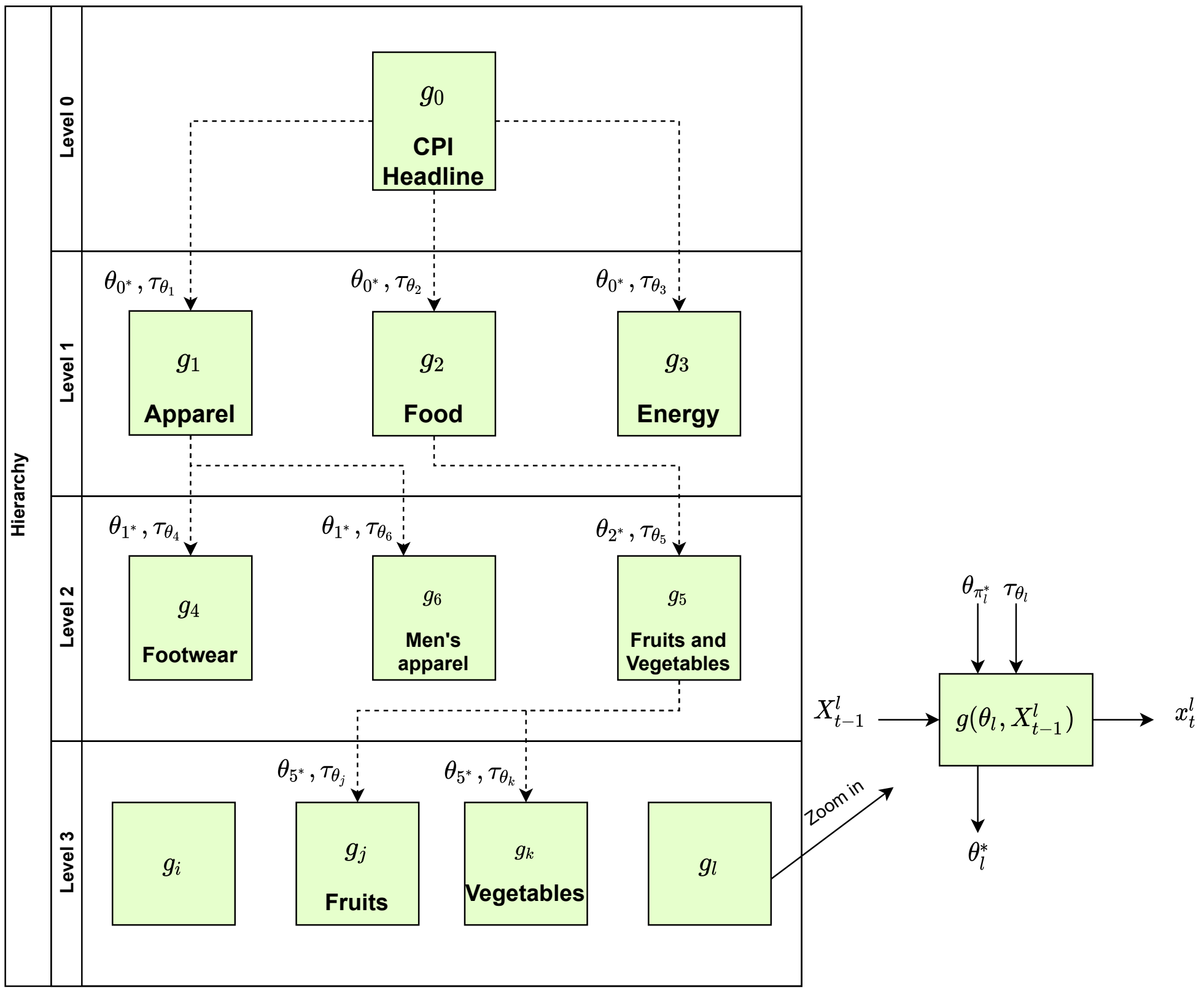

Let us now denote the aggregation of all series from all levels by $X = \{X^n_{T_n}\}_{n \in \mathcal{I}}$. Similarly, we denote by $\theta = \{\theta_n\}_{n \in \mathcal{I}}$ and $\mathrm{T} = \{\tau_n\}_{n \in \mathcal{I}}$ the aggregation of all the RNN parameters and precision parameters from all levels, respectively. Note that $X$ (the data) is observed, $\theta$ are unobserved *learned* variables, and T are determined by Equation (6). The hyper-parameter $\alpha$ from Equation (6) is set by a cross-validation procedure.

With these definitions at hand, we now proceed with the Bayes rule. From Equation 4 and Equation (5), we extract the posterior probability:

$$p(\theta|X,\mathrm{T}) = \frac{p(X|\theta,\mathrm{T})p(\theta)}{P(X)} \propto \prod_{n \in \mathcal{I}} \prod_{t=1}^{T_n} \mathcal{N}(x_t^n; g(\theta_n, X_{t-1}^n), \tau_n^{-1}) \prod_{n \in \mathcal{I}} \mathcal{N}(\theta_n; \theta_{\pi_n}, \tau_{\theta_n}^{-1}\mathbf{I}). \tag{7}$$

HRNN optimization follows a *Maximum A-Posteriori* (MAP) approach. Namely, we wish to find optimal parameter values $\theta^*$ such that:

$$\theta^* = \underset{\theta}{\operatorname{argmax}} \log p(\theta|X,\mathrm{T}). \tag{8}$$

Note that the objective in Equation (8) depends on the parametric function $g$. HRNN is a general framework that can use any RNN, e.g., Simple RNN, LSTM, GRU, etc. In this work, we chose $g$ to be a scalar GRU because GRUs are capable of long-term memory but with fewer parameters than LSTMs. Hence, each node $n$ is associated with a GRU with its own parameters: $\theta_n = [u_n^z, u_n^r, u_n^v, w_n^z, w_n^r, w_n^v, b_n^z, b_n^r, b_n^v]$. Then, $g(\theta_n, X_t^n)$ is computed by $t$ successive applications of the GRU to $x_i^n$ with $1 \leq i \leq t$ according to Equation (3). Finally, the HRNN optimization proceeds with stochastic gradient ascent over the objective in Equation (8). Figure 4 depicts an illustration of the entire HRNN architecture.

### *4.2. HRNN Inference*

In machine learning, after the model's parameters have been estimated in the *training* process, it can be applied to make predictions in a process known as *inference*. In our case, equipped with the MAP estimate $\theta^*$, inference with the HRNN model is achieved as follows: Given a sequence of historical CPI values $X_t^n$ for node $n$, we predict the next CPI value $y_{t+1}^n = g(\theta_n, X_t^n)$, as explained in Section 4.1. This type of prediction is for next month's CPI, namely, horizon $h = 0$. In this work, we also test the ability of the model to perform predictions for further horizons $h \in \{0, .., 8\}$. The $h$-horizon predictions are obtained in a recursive manner, whereby each predicted value $y_t^n$ is fed back as an input for the prediction of $y_{t+1}^n$. As expected, Section 6 shows that forecasting accuracy gradually degrades as horizon $h$ increases.

## 5. Dataset

This work is based on monthly CPI data released by the US Bureau of Labor and Statistics (BLS). In what follows, we discuss the dataset's characteristics and our pre-processing procedures. For the sake of reproducibility, the final version of the processed data is available in our HRNN code.

### *5.1. The US Consumer Price Index*

The official CPI of each month is released by the BLS several days into the following month. The price tags are collected in 75 urban areas throughout the US from about 24,000 retail and service establishments. The housing and rent rates are collected from about 50,000 landlords and tenants across the country. The BLS releases two different measurements according to urban demographics:

1. The **CPI-U** represents the CPI for urban consumers and covers approximately 93% of the total population. According to the Consumer Expenditure Survey, the CPI items and their relative weights are derived from their estimated expenditure. These items and their weights are updated each year in January.

2. The **CPI-W** represents the CPI for urban wage earners and clerical workers and covers about 29% of the population. This index is focused on households with at least 50 percent of income coming from clerical or wage-paying jobs, and at least one of the household's earners must have been employed for at least 70% of the year. CPI-W indicates changes in the cost of benefits, as well as future contract obligations.

In this work, we focus on CPI-U, as it is generally considered the best measure for the average cost of living in the US. Monthly CPI-U data per product is generally available from January 1994. Our samples thus span from January 1994 to March 2019. Note that throughout the years, new indexes were added, and some indexes have been omitted. Consequently, hierarchies can change, which contributes to the challenge of our exercise.

### 5.2. *The CPI Hierarchy*

The CPI-U is an eight-level deep hierarchy comprising 424 different nodes (indexes). Level 0 represents the headline CPI, or the aggregated index of all components. An index at any level is associated with a weight between 0-100, which represents its contribution to the headline CPI at level 0. Level 1 consists of the 8 main aggregated categories or sectors: (1) “Food and Beverages”, (2) “Housing”, (3) “Apparel”, (4) “Transportation”, (5) “Medical Care”, (6) “Recreation”, (7) “Education and Communication”, and (8) “Other Goods and Services”. Mid-levels (2-5) consist of more specific aggregations e.g., “Energy Commodities”, “Household Insurance”, etc. The lower levels (6-8) consists of fine-grained indexes, e.g., “Apples”, “Bacon and Related Products”, “Eyeglasses and Eye Care”, “Tires”, “Airline fares”, etc. Tables 7 and 8 (in Appendix A) depict the first three hierarchies of the CPI (levels 0-2).

### 5.3. *Data Preparation*

We used publicly available data from the BLS website[2]. However, the BLS releases hierarchical data on a monthly basis in separate files. Hence, separate monthly files from January 1994 until March 2019 were processed and aggregated to create a single repository. Moreover, the format of these files has changed over the years (e.g., txt, pdf, and csv formats were all in use) and a significant effort was made in order to parse the changing formats from different time periods.

The hierarchical CPI data is released in terms of monthly index values. We transformed the CPI values to monthly logarithmic change rates as follows: We denote by $x_t$ the CPI value (of any node) at month $t$. The logarithmic change rate at month $t$ is denoted by $rate(t)$ and given by:

$$rate(t) = 100 \times \log\left(\frac{x_t}{x_{t-1}}\right). \tag{9}$$

Unless otherwise mentioned, the remainder of the paper relates to monthly logarithmic change rates as in Equation (9).

We split the data into a *training* dataset and a *test* dataset as follows: For each time series, we kept the first (early in time) 70% of the measurements for the *training* dataset. The remaining 30% of the measurements were removed from the *training* dataset and used to form the *test* dataset. The *training* dataset was used to train the HRNN model as well as the other baselines. The *test* dataset was used for evaluations. The results in Section 6 are based on this split.

Table 1 summarizes the number of data points and general statistics of the CPI time series after applying Equation (9). When comparing the headline CPI with the full

[2] www.bls.gov/cpi

Table 1: Descriptive Statistics

| Data set | # Monthly Measurements | Mean | STD | Min | Max | # of Indexes | Avg. Measurements per Index |
|---|---|---|---|---|---|---|---|
| Headline Only | 303 | 0.18 | 0.33 | -1.93 | 1.22 | 1 | 303 |
| Level 1 | 6742 | 0.17 | 0.96 | -18.61 | 11.32 | 34 | 198.29 |
| Level 2 | 6879 | 0.12 | 1.10 | -19.60 | 16.81 | 46 | 149.54 |
| Level 3 | 7885 | 0.17 | 1.31 | -34.23 | 16.37 | 51 | 121.31 |
| Level 4 | 7403 | 0.08 | 1.97 | -35.00 | 28.17 | 58 | 107.89 |
| Level 5 | 10809 | 0.01 | 1.43 | -21.04 | 242.50 | 92 | 87.90 |
| Level 6 | 7752 | 0.09 | 1.49 | -11.71 | 16.52 | 85 | 86.13 |
| Level 7 | 4037 | 0.11 | 1.53 | -11.90 | 9.45 | 50 | 80.74 |
| Level 8 | 595 | 0.08 | 1.56 | -5.27 | 5.02 | 7 | 85.00 |
| Full Hierarchy | 52405 | 0.10 | 1.75 | -35.00 | 242.50 | 424 | 123.31 |

*Notes:* General statistics of the headline CPI and CPI-U for each level in the hierarchy and the full hierarchy of indexes.

hierarchy, we see that at lower levels the standard deviation (STD) is significantly higher and the dynamic range is larger, implying much more volatility. The average number of measurements per index decreases at the lower levels of the hierarchy as not all indexes are available for the entire period.

Figure 5 depicts box plots of the CPI change rate distributions at different levels. The boxes depict the median value and the upper 75'th and lower 25'th percentiles. The whiskers indicate the overall minimum and maximum rates. Figure 5 further emphasizes that the change rates are more volatile as we go down the CPI hierarchy.

High dynamic range, high standard deviation, and less training data are all indicators of the difficulty of making predictions inside the hierarchy. Based on this information, we can expect that the disaggregated component predictions inside the hierarchy will be more difficult than the headline.

Finally, Figure 6 depicts a box plot of the CPI change rate distribution for different sectors. We notice that some sectors (e.g., apparel and energy) suffer from higher volatility than others. As expected, predictions for these sectors will be more difficult.

## 6. Evaluation and Results

We evaluate HRNN and compare it with well-known baselines for inflation prediction as well as some alternative machine learning approaches. We use the following notation: Let $x_t$ be the CPI log-change rate at month $t$. We consider models for $\hat{x}_t$ - an estimate for $x_t$ based on historical values. Additionally, we denote by $\varepsilon_t$ the estimation error at time $t$. In all cases, the $h$-horizon forecasts were generated by recursively iterating the one-step forecasts forward. Hyper-parameters were set through a 10-fold cross-validation procedure.

### 6.1. *Baseline Models*

We compare HRNN with the following CPI prediction baselines:

1. **Autoregression (AR) -** The AR($\rho$) estimates $\hat{x}_t$ based on the previous $\rho$ months as follows: $\hat{x}_t = \alpha_0 + \left(\sum_{i=1}^{\rho} \alpha_i x_{t-i}\right) + \varepsilon_t$, where $\{\alpha_i\}_{i=0}^{\rho}$ are the model's parameters.

**Figure 5.** *Box plots of monthly inflation rate per hierarchy level.*

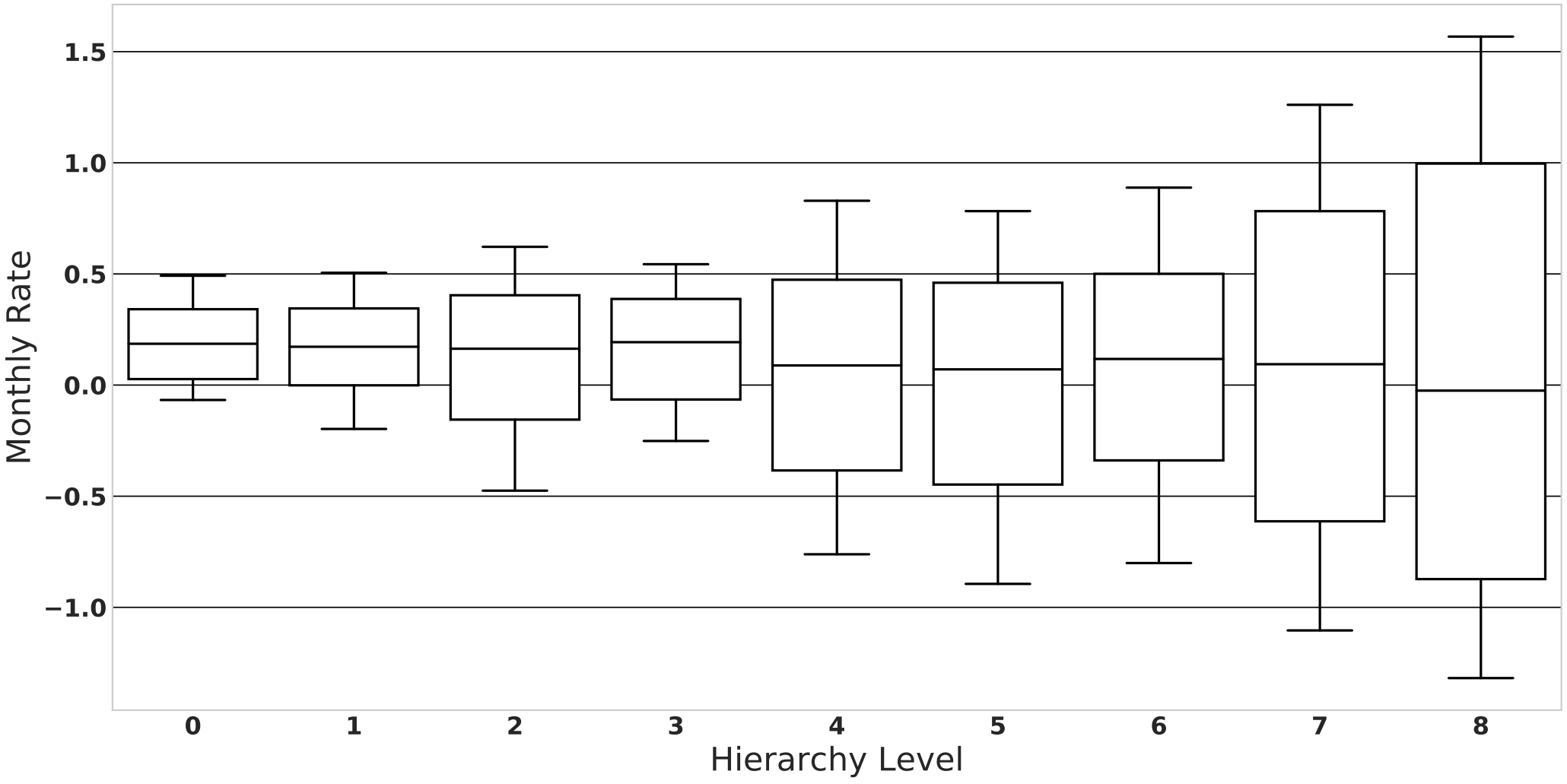

**Figure 6.** *Box plots of monthly inflation rate per sector.*

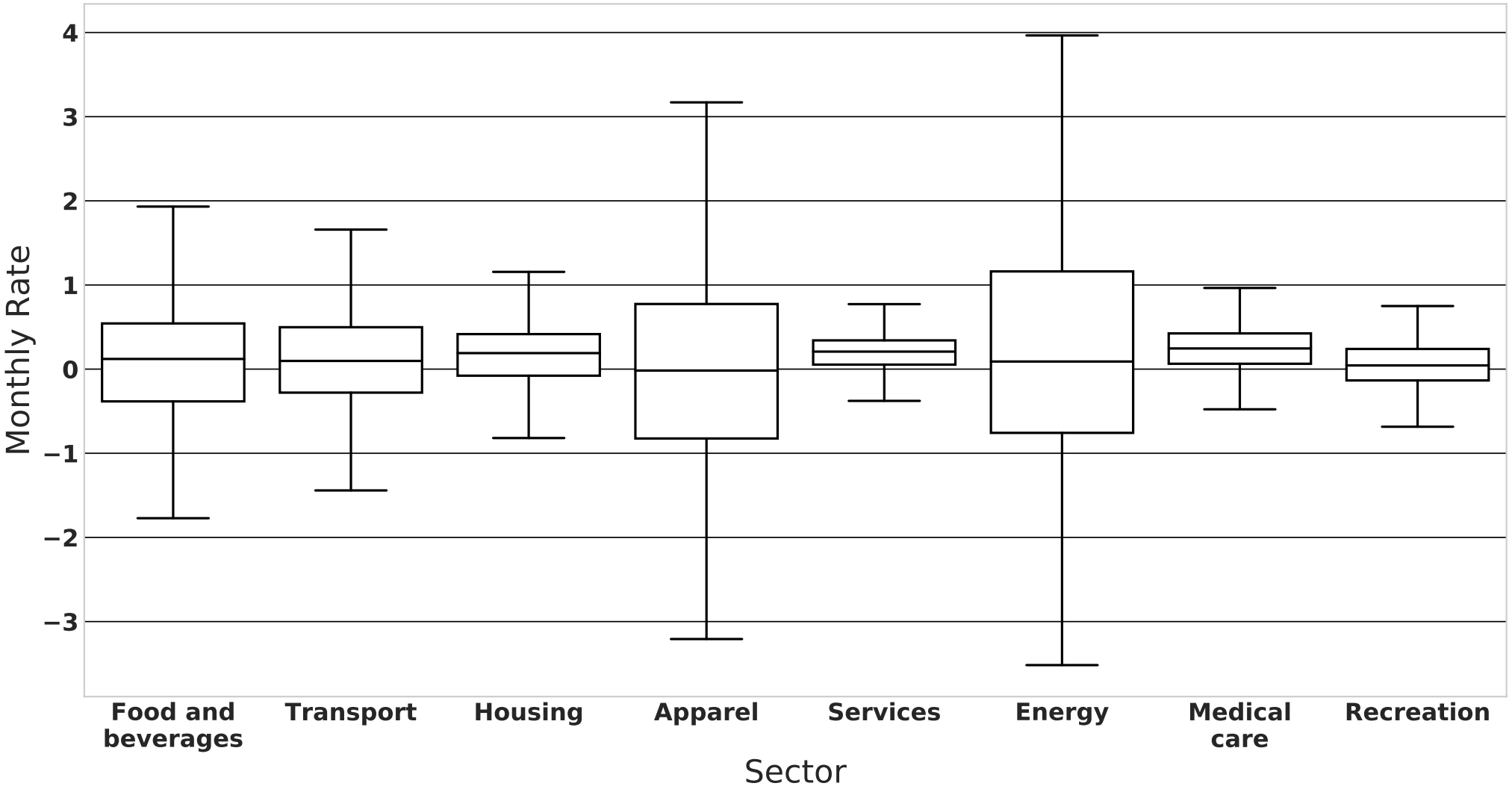

2. **Phillips Curve (PC) -** A PC($\rho$) is an extension of AR($\rho$) that considers the unemployment rate $u_t$ at month $t$ in CPI forecasting model such as: $\hat{x}_t = \alpha_0 + \left(\sum_{i=1}^{\rho} \alpha_i x_{t-i}\right) + \beta u_{t-1} + \varepsilon_t$, where $\{\alpha_i\}_{i=0}^{\rho}$ and $\beta$ are the model's parameters.

3. **Vector Autoregression (VAR) -** The VAR($\rho$) model is a multivariate generalization of AR($\rho$). It is frequently used to model two or more time series together. VAR($\rho$) estimates next month's values of $k$ time series based on their historical values from the previous $\rho$ months as follows: $\hat{X}_t = A_0 + (\sum_{i=1}^{\rho} A_i X_{t-i}) + \epsilon_t$, where $X_t$ are the last $\rho$ values from $k$ different time series at month $t$, and $\hat{X}_t$ are the model's estimates of these values, $\{A_i\}_{i=0}^{\rho}$ are a $(k \times k)$ matrices of parameters, and $\epsilon_t$ is a vector of error terms.

4. **Random Walk (RW) -** We consider the RW($\rho$) model of Atkeson et al. (2001). RW($\rho$) is a simple, yet effective, model that predicts next month's CPI as an average of the last $\rho$ months by: $\hat{y}_t = \frac{1}{\rho}\sum_{i=1}^{\rho} x_{t-i} + \varepsilon_t$.

5. **Auto Regression in Gap (AR-GAP) -** The AR-GAP model subtracts a fixed inflation trend before predicting the inflation in gap (Faust and Wright, 2013). Inflation gap is defined as $g_t = x_t - \tau_t$, where $\tau_t$ is the inflation trend at time $t$ which represents a slowly-varying local mean. This trend value is estimated using RW($\rho$) as follows: $\tau_t = \frac{1}{\rho}\sum_{i=1}^{\rho} x_{t-i}$. By accounting for the local inflation trend $\tau_t$, the model attempts to increase stationarity in $g_t$ and estimate it by $\hat{g}_t = \alpha_0 + \sum_{i=1}^{\rho} \alpha_i g_{t-i} + \varepsilon_t$, where $\{\alpha_i\}_{i=0}^{\rho}$ are the model's parameters. Finally, $\tau_t$ is added back to $\hat{g}_t$ to achieve the forecast for the final inflation prediction: $\hat{x}_t = \hat{g}_t + \tau_t$.

6. **Logistic Smooth Transition Auto Regressive Model (LSTAR) -** The LSTAR is an extension of AR that allows for changes in the model parameters according to a transition variable $F(t; c, \gamma)$. LSTAR($\rho, c, \gamma$) consists of two AR($\rho$) components that describe two trends in the data (high and low), and a nonlinear transition function that links them as follows:

$$\hat{x}_t = \left(\alpha_0 + \sum_{i=1}^{\rho} \alpha_i x_{t-i}\right)(1 - F(t; \gamma, c)) + \left(\beta_0 + \sum_{i=1}^{\rho} \beta_i x_{t-i}\right) F(t; \gamma, c) + \varepsilon_t, \tag{10}$$

where $F(t; \gamma, c) = \frac{1}{1+e^{-\gamma(t-c)}}$ is a first-order logistic transition function that depends on the location parameter $c$, and a smoothing parameter $\gamma$. The location parameter $c$ can be interpreted as the threshold between the two AR($\rho$) regimes, in the sense that the logistic function changes monotonically from 0 to 1 as $t$ increases and balances symmetrically at $t = c$ (van Dijk et al., 2002). The model's parameters are $\{\alpha_i\}_{i=0}^{\rho}$ and $\{\beta_i\}_{i=0}^{\rho}$, while $\gamma$, and $c$ are hyper-parameters.

7. **Random Forests (RF) -** The RF($\rho$) model is an ensemble learning method which builds a set of decision trees (Song and Ying, 2015) in order to mitigate overfitting and improve generalization (Breiman, 2001). At prediction time, the average prediction of the individual trees is returned. The inputs to the RF($\rho$) model are the last $\rho$ samples and the output is the predicted value for the next month.

8. **Gradient Boosted Trees (GBT) -** The GBT($\rho$) model (Friedman, 2002) is based on an ensemble of decision trees which are trained in a stage-wise fashion similar to other boosting models (Schapire, 1999). Unlike RF($\rho$) which averages the prediction of several decision trees, the GBT($\rho$) trains each tree to minimize the remaining residual error of all previous trees. At prediction time, the sum of predictions of all the trees is returned. The inputs to the GBT($\rho$) model are the last $\rho$ samples and the output is the predicted value for the next month.

9. **Fully Connected Neural Network (FC) -** The FC($\rho$) model is a fully connected neural network with one hidden layer and a ReLU activation (Ramachandran et al., 2017). The output layer employs no activation to formulate a regression problem with a squared loss optimization. The inputs to the FC($\rho$) model are the last $\rho$ samples and the output is the predicted value for the next month.

10. **Deep Neural Network (Deep-NN) -** The Deep-NN($\rho$) model is a deep neural network consisting of 10 layers with 100 neurons as in Olson et al. (2018), which was shown to perform well for inflation prediction (Goulet Coulombe, 2020). We used the original set-up of Olson et al. (2018) and tuned its hyper-parameters as follows: learning rate was set to $lr = 0.005$, training lasted 50 epochs (instead of 200), and the ELU activation functions (Clevert et al., 2016) were replaced by ReLU activation functions. These changes yielded more accurate predictions, hence we decided to include them in all our evaluations. The inputs to the Deep-NN($\rho$) model are the last $\rho$ samples and the output is the predicted value for the next month.

11. **Deep Neural Network with Unemployment (Deep-NN + Unemployment) -** Similar to PC($\rho$) which extends AR($\rho$) by including unemployment data, the Deep-NN($\rho$) + Unemployment model extends Deep-NN($\rho$) by including the last $\rho$ samples of the unemployment rate $u_t$. In terms of hyper-parameters, we used identical values as in the Deep-NN($\rho$).

### 6.2. *Ablation Models*

In order to demonstrate the contribution of hierarchical component of the HRNN model, we conducted an ablation study that considered "simpler" alternatives to HRNN based on GRUs without the hierarchical component:

1. **Single (S-GRU) -** The S-GRU($\rho$) is a single GRU unit that receives the last $\rho$ values as inputs in order to predict the next value. In GRU($\rho$), a single GRU is used for all the time series that comprise the CPI hierarchy. This baseline utilizes all the benefits of a GRU but assumes that the different components of the CPI behave similarly and a single unit is sufficient to model all the nodes.

2. **Independent GRUs (I-GRUs) -** In I-GRUs($\rho$), we trained a different GRU($\rho$) unit for each CPI node. The S-GRU and I-GRU approaches represent two extremes: The first attempts to model all the CPI nodes with a single model, while the second treats each node separately. I-GRUs($\rho$) is equivalent to a variant of HRNN that ignores the hierarchy by setting the precision parameter $\tau_{\theta_n} = 0; \forall n \in \mathcal{I}$. Namely, this is a simple variant of HRNN that trains independent GRUs, one for each index in the hierarchy.

3. **K-Nearest Neighbors GRU (KNN-GRU) -** In order to demonstrate the contribution of the hierarchical structure of HRNN, we devised the KNN-GRU($\rho$) baseline. KNN-GRU attempts to utilize information from multiple Pearson-correlated CPI nodes without employing the hierarchical informative priors. Hence, KNN-GRU presents a "simpler" alternative to HRNN that replaces the hierarchical structure with elementary vector GRUs as follows: First, the $k$ nearest neighbors of each CPI node were found using the Pearson correlation measure. Then, separate vector GRU($\rho$) units were trained for each CPI aggregate along its $k$ most similar nodes using the last $\rho$ values of node $n$ and its $k$-nearest nodes. By doing so, the KNN-GRU($\rho$) baseline was able to utilize both the benefits of GRU units together with relevant information that comes from correlated nodes.

### 6.3. *Evaluation Metrics*

Following Faust and Wright (2013) and Aparicio and Bertolotto (2020), we report results in terms of three evaluation metrics:

1. **Root Mean Squared Error (RMSE) -** The RMSE is given by:

$$RMSE = \sqrt{\frac{1}{T}\sum_{t=1}^{T}(x_t - \hat{x}_t)^2}, \tag{11}$$

where $x_t$ are the monthly change rate for month $t$, and $\hat{x}_t$ are the corresponding predictions.

2. **Pearson Correlation Coefficient -** The Pearson correlation coefficient $\phi$ is given by:

$$\phi = \frac{COV(X_T, \hat{X}_T)}{\sigma_X \times \sigma_{\hat{X}}}, \tag{12}$$

where $COV(X_T, \hat{X}_T)$ is the covariance between the series of actual values and the predictions, and $\sigma_{X_T}$ and $\sigma_{\hat{X}_T}$ are the standard deviations of the actual values and the predictions, respectively.

3. **Distance Correlation Coefficient -** In contrast to the Pearson correlation measure, which detects linear associations between two random variables, the distance correlation measure can also detect nonlinear correlations (Székely et al., 2007; Zhou, 2012). The distance correlation coefficient $r_d$ is given by:

$$r_d = \frac{\mathrm{dCov}(X_T, \hat{X}_T)}{\sqrt{\mathrm{dVar}(X_T) \times \mathrm{dVar}(\hat{X}_T)}}, \tag{13}$$

where $\mathrm{dCov}(X_T, \hat{X}_T)$ is the distance covariance between the series of actual values and the predictions, and $\mathrm{dVar}(X_T)$ and $\mathrm{dVar}(\hat{X}_T)$ are the distance variance of the actual values and the predictions, respectively.

### 6.4. *Results*

The HRNN model is unique in its ability to utilize information from higher levels in the CPI hierarchy in order to make predictions at lower levels. Therefore, we provide results for each level of the CPI hierarchy - overall 424 disaggregated indexes belonging to 8 different hierarchies. For the sake of completion, we also provide results for the headline CPI index by itself. It is important to note that in this case, the HRNN model cannot utilize its hierarchical mechanism and has no advantage over the alternatives, so we do not expect it to outperform.

Table 2: Average Results on Disaggregated CPI Components

| Model Name | RMSE per horizon AR(1)=1.00 | | | | | | Correlation (at horizon=0) | |
|---|---|---|---|---|---|---|---|---|
| | 0 | 1 | 2 | 3 | 4 | 8 | Pearson | Distance |
| AR(1) | 1.00 | 1.00 | 1.00 | 1.00 | 1.00 | 1.00 | 0.06 | 0.05 |
| AR(2) | 1.00 | 1.00 | 1.00 | 1.00 | 1.00 | 1.00 | 0.08 | 0.06 |
| AR(3) | 1.00 | 1.00 | 1.00 | 1.00 | 1.00 | 1.00 | 0.08 | 0.06 |
| AR(4) | 1.00 | 1.00 | 1.00 | 1.00 | 1.00 | 1.00 | 0.09 | 0.07 |
| AR-GAP(3) | 1.00 | 1.00 | 1.00 | 1.00 | 1.00 | 1.00 | 0.08 | 0.06 |
| AR-GAP(4) | 1.00 | 1.00 | 1.00 | 1.00 | 1.00 | 1.00 | 0.09 | 0.07 |
| RW(4) | 1.00 | 1.00 | 1.00 | 1.00 | 1.00 | 1.00 | -0.05 | -0.04 |
| Phillips(4) | 1.00 | 1.00 | 1.00 | 1.00 | 0.98 | 1.00 | -0.05 | -0.04 |
| VAR(1) | 1.03 | 1.03 | 1.04 | 1.03 | 1.04 | 1.05 | 0.04 | 0.03 |
| VAR(2) | 1.03 | 1.03 | 1.04 | 1.03 | 1.04 | 1.05 | 0.06 | 0.03 |
| VAR(3) | 1.03 | 1.03 | 1.03 | 1.03 | 1.04 | 1.05 | 0.06 | 0.03 |
| VAR(4) | 1.02 | 1.03 | 1.03 | 1.03 | 1.03 | 1.04 | 0.07 | 0.04 |
| LSTAR($\rho = 4, c = 2, \gamma = 0.3$) | 1.04 | 1.07 | 1.07 | 1.07 | 1.08 | 1.1 | 0.09 | 0.07 |
| GBT(4) | 0.83 | 0.83 | 0.83 | 0.84 | 0.84 | 0.86 | 0.18 | 0.27 |
| RF(4) | 0.84 | 0.85 | 0.86 | 0.86 | 0.86 | 0.87 | 0.19 | 0.29 |
| FC(4) | 1.03 | 1.03 | 1.04 | 1.04 | 1.04 | 1.05 | 0.12 | 0.09 |
| Deep-NN(4) | 0.90 | 0.90 | 0.90 | 0.90 | 0.91 | 0.91 | 0.13 | 0.22 |
| Deep-NN(4) + Unemployment | 0.85 | 0.85 | 0.85 | 0.85 | 0.85 | 0.86 | 0.12 | 0.22 |
| S-GRU(4) | 1.02 | 1.06 | 1.06 | 1.07 | 1.04 | 1.12 | 0.10 | 0.08 |
| I-GRU(4) | 0.83 | 0.84 | 0.85 | 0.85 | 0.86 | 0.89 | 0.17 | 0.13 |
| KNN-GRU(1) | 0.91 | 0.93 | 0.96 | 0.97 | 0.96 | 0.96 | 0.19 | 0.15 |
| KNN-GRU(2) | 0.90 | 0.93 | 0.95 | 0.97 | 0.96 | 0.96 | 0.20 | 0.15 |
| KNN-GRU(3) | 0.89 | 0.92 | 0.95 | 0.96 | 0.96 | 0.95 | 0.20 | 0.15 |
| KNN-GRU(4) | 0.89 | 0.91 | 0.95 | 0.95 | 0.95 | 0.95 | 0.20 | 0.15 |
| HRNN(1) | 0.79 | 0.79 | 0.81 | 0.81 | 0.81 | 0.83 | 0.23 | 0.28 |
| HRNN(2) | **0.78** | 0.79 | 0.81 | 0.81 | 0.80 | 0.82 | 0.22 | 0.29 |
| HRNN(3) | 0.79 | **0.78** | 0.80 | 0.81 | 0.81 | 0.81 | 0.23 | **0.30** |
| **HRNN(4)** | **0.78** | **0.78** | **0.79** | **0.79** | **0.79** | **0.80** | **0.24** | 0.29 |

*Notes:* Average results across all 424 inflation indexes that make up the headline CPI. The RMSE results are relative to the $AR(1)$ model and normalized according to its results, i.e., $\frac{RMSE_{Model}}{RMSE_{AR(1)}}$. Results are statistically significant according to Diebold-Mariano test with $p < 0.02$.

Table 2 depicts the average results from all the disaggregated indexes in the CPI hierarchy. We present prediction results for horizons 0, 1, 2, 3, 4, and 8 months. The results are relative to the $AR(1)$ model and normalized according to: $\frac{RMSE_{Model}}{RMSE_{AR(1)}}$. In HRNN we set $\alpha = 1.5$, and the V-GRU($\rho$) models were based on $k = 5$ nearest neighbors. Table 2 shows that different versions of the HRNN model repeatedly outperform the alternatives at any horizon. Notably, HRNN is superior to I-GRU, which emphasizes the importance of using hierarchical information and the superiority of HRNN over regular GRUs. Additionally, the HRNN is also superior to the different KNN-GRU models, which emphasizes the specific way HRNN employs informative priors based on the CPI hierarchy. These results are statistically significant according to Diebold and Mariano (1995) pairwise tests for a squared loss-differential with p-values below 0.02. Additionally, we performed a Model Confidence Set (MCS) test (Hansen et al., 2011) for the leading models: RF(4), Deep-NN(4), Deep-NN(4) + Unemployment, GBT(4), IGRU(4), HRNN(1), HRNN(2), HRNN(3), and HRNN(4). MCS removed all the

Table 3: CPI Headline Only

| Model Name* | RMSE per horizon AR(1)=1.00 | | | | | | Correlation (at horizon=0) | |
|---|---|---|---|---|---|---|---|---|
| | 0 | 1 | 2 | 3 | 4 | 8 | Pearson | Distance |
| AR(1) | 1.00 | 1.00 | 1.00 | 1.00 | 1.00 | 1.00 | 0.29 | 0.22 |
| AR(2) | 1.00 | 0.97 | 0.99 | 1.01 | 1.00 | 0.98 | 0.32 | 0.24 |
| AR(3) | 1.00 | 0.98 | 0.98 | 1.00 | 0.96 | 0.97 | 0.33 | 0.25 |
| AR(4) | 1.00 | 0.95 | 0.95 | 0.96 | 0.93 | 0.96 | 0.33 | 0.25 |
| AR-GAP(3) | 1.00 | 0.98 | 0.98 | 1.00 | 0.96 | 0.97 | 0.33 | 0.25 |
| AR-GAP(4) | 0.99 | 0.95 | 0.95 | 0.96 | 0.93 | 0.96 | 0.33 | 0.25 |
| RW(4) | 1.05 | 0.98 | 0.99 | 1.01 | 0.97 | 0.96 | -0.08 | -0.06 |
| Phillips(4) | 0.93 | 0.94 | 0.95 | 0.95 | 0.93 | 0.95 | 0.33 | 0.25 |
| LSTAR($\rho = 4, c = 2, \gamma = 0.3$) | 0.98 | 0.95 | 0.95 | 0.97 | 0.95 | 0.95 | 0.32 | 0.24 |
| RF(4) | 1.05 | 1.06 | 1.03 | 1.07 | 1.04 | 1.03 | 0.27 | 0.28 |
| GBT(4) | 0.97 | 0.99 | 0.93 | 0.95 | 0.93 | 0.93 | 0.25 | 0.35 |
| FC(4) | **0.92** | **0.94** | 0.94 | 0.96 | 0.93 | 0.94 | 0.33 | 0.25 |
| Deep-NN(4) | 0.94 | 0.97 | 0.96 | 0.98 | 0.94 | 0.92 | 0.31 | 0.32 |
| Deep-NN(4) + Unemployment | 1.00 | 0.97 | **0.92** | **0.94** | **0.92** | **0.91** | **0.37** | 0.32 |
| HRNN(4) / GRU(4) | 1.00 | 0.97 | 0.99 | 0.99 | 0.96 | 0.99 | 0.35 | **0.37** |

*Notes:* Prediction results for the CPI headline index alone. The RMSE results are relative to the $AR(1)$ model and normalized according to its results, i.e., $\frac{RMSE_{Model}}{RMSE_{AR(1)}}$.

baselines and left only the four HRNN variants, with HRNN(4) as the leading model ($p_{HRNN(4)} = 1.00$).

For the sake of completion, we also provide results for predictions at the head of the CPI index. Table 3 summarizes these results. When considering only the headline, the hierarchical mechanism of HRNN is redundant and the model is identical to a single GRU unit. In this case, we do not observe much advantage for employing the HRNN model. In contrast, we see an advantage for the other deep learning models such as FC(4) and Deep-NN(4) + Unemployment that outperform the more "traditional" approaches.

Table 4: HRNN(4) vs. I-GRU(4) at different levels of the CPI hierarchy with respect to AR(1)

| Hierarchy Level | HRNN(4) RMSE per horizon AR(1)=1.00 | | | | HRNN(4) Correlation (at horizon=0) | | I-GRU(4) RMSE per horizon AR(1)=1.00 | | | | I-GRU(4) Correlation (at horizon=0) | |
|---|---|---|---|---|---|---|---|---|---|---|---|---|
| | 0 | 2 | 4 | 8 | Pearson | Distance | 0 | 2 | 4 | 8 | Pearson | Distance |
| Level 1 | 0.95 | 0.97 | 0.99 | 1.00 | 0.33 | 0.37 | 0.98 | 0.98 | 0.99 | 0.97 | 0.25 | 0.38 |
| Level 2 | 0.91 | 0.90 | 0.91 | 0.91 | 0.30 | 0.35 | 0.90 | 092 | 0.94 | 0.93 | 0.26 | 0.34 |
| Level 3 | 0.79 | 0.79 | 0.80 | 0.81 | 0.21 | 0.31 | 0.82 | 0.89 | 0.94 | 0.94 | 0.23 | 0.37 |
| Level 4 | 0.77 | 0.77 | 0.76 | 0.77 | 0.26 | 0.32 | 0.84 | 0.87 | 0.90 | 0.92 | 0.20 | 0.33 |
| Level 5 | 0.79 | 0.77 | 0.77 | 0.80 | 0.21 | 0.31 | 0.85 | 0.89 | 0.89 | 0.93 | 0.22 | 0.29 |
| Level 6 | 0.75 | 0.76 | 0.81 | 0.81 | 0.19 | 0.23 | 0.85 | 0.89 | 0.90 | 0.92 | 0.21 | 0.21 |
| Level 7 | 0.75 | 0.78 | 0.77 | 0.80 | 0.17 | 0.17 | 0.87 | 0.89 | 0.92 | 0.94 | 0.18 | 0.15 |
| Level 8 | 0.72 | 0.78 | 0.77 | 0.78 | 0.10 | 0.23 | 0.89 | 0.90 | 0.92 | 0.94 | 0.10 | 0.12 |

*Notes:* The RMSE results are relative to the $AR(1)$ model and normalized according to its results, i.e., $\frac{RMSE_{Model}}{RMSE_{AR(1)}}$.

Table 5: HRNN(4) vs. I-GRU(4) results for different CPI sectors with respect to AR(1)

| Industry Sector | HRNN(4) | | | | | | I-GRU(4) | | | | | |
|---|---|---|---|---|---|---|---|---|---|---|---|---|
| | RMSE per horizon AR(1)=1.00 | | | | Correlation (at horizon=0) | | RMSE per horizon AR(1)=1.00 | | | | Correlation (at horizon=0) | |
| | 0 | 2 | 4 | 8 | Pearson | Distance | 0 | 2 | 4 | 8 | Pearson | Distance |
| **Apparel** | 0.83 | 0.87 | 0.84 | 0.88 | 0.04 | 0.19 | 0.88 | 0.88 | 0.85 | 0.92 | 0.05 | 0.23 |
| **Energy** | 0.94 | 0.96 | 0.99 | 0.98 | 0.34 | 0.32 | 0.94 | 0.98 | 1.02 | 0.99 | 0.18 | 0.28 |
| **Food & beverages** | 0.72 | 0.73 | 0.75 | 0.76 | 0.22 | 0.13 | 0.80 | 0.80 | 0.81 | 0.82 | 0.18 | 0.22 |
| **Housing** | 0.79 | 0.80 | 0.82 | 0.82 | 0.17 | 0.24 | 0.77 | 0.79 | 0.82 | 0.82 | 0.18 | 0.27 |
| **Medical care** | 0.79 | 0.82 | 0.81 | 0.82 | 0.03 | 0.17 | 0.79 | 0.83 | 0.83 | 0.84 | 0.08 | 0.15 |
| **Recreation** | 0.99 | 0.99 | 1.00 | 1.00 | 0.05 | 0.17 | 1.00 | 0.99 | 1.00 | 1.00 | -0.07 | 0.17 |
| **Services** | 0.90 | 0.92 | 0.95 | 0.94 | 0.04 | 0.15 | 0.89 | 0.94 | 0.95 | 0.96 | 0.02 | 0.21 |
| **Transportation** | 0.83 | 0.84 | 0.85 | 0.85 | 0.27 | 0.28 | 0.82 | 0.85 | 0.86 | 0.88 | 0.26 | 0.36 |

*Notes:* The RMSE results are relative to the $AR(1)$ model and normalized according to its results, i.e., $\frac{RMSE_{Model}}{RMSE_{AR(1)}}$.

Table 4 depicts the results of HRNN(4), the best model, across all hierarchies (1-8, excluding the headline). Additionally, we included the results of the best ablation model, the I-GRU(4) model, for comparison. Results are averaged over all disaggregated components and normalized by the AR(1) model RMSE as before. As evident from Table 4, the HRNN model shows the best relative performance at the lower levels of the hierarchy where the CPI indexes are more volatile and the hierarchical priors are most effective.

Table 5 compares the results of HRNN(4) across different sectors. Again, we included the results of the I-GRU(4) model for comparison. The results are averaged over all disaggregated components and presented as normalized gains with respect to the AR(1) model as before. The best relative improvement of HRNN(4) model appears to be in the Food and Beverages group. This can be explained by the fact that the Food and Beverages sub-hierarchy is the deepest and most elaborate hierarchy in the CPI tree. When the hierarchy is deeper and more elaborate, HRNN advantages are emphasized.

Finally, Figure 7 depicts specific examples of three disaggregated indexes: Tomatoes, Bread, and Information Technology. The solid red line presents the actual CPI values. The dashed green line presents HRNN(4) predictions, while the dotted blue line presents the I-GRU(4) predictions. These indexes are located down at the bottom of the CPI hierarchy and suffer from relatively high volatility. The HRNN(4) model seems to track and predict the trends of the real index accurately and often perform better then I-GRU(4). As can be seen, I-GRU's predictions appear to be more "conservative" than HRNN. At first, this may appear counterintuitive, as HRNN has more regularization than I-GRU. However, this additional regularization is actually *informative regularization* coming from the parameters of the upper levels in the CPI hierarchy which allows the HRNN model to be more expressive without overfitting. In contrast, in order to ensure that I-GRU does not overfit the training data, its other regularization techniques such the learning rate hyper-parameter and the early stopping procedure prevent the I-GRU model from becoming overconfident. Figure 9 and Figure 10 in Appendix A depict additional examples for a large variety of disaggregated CPI components.

**Figure 7.** *Examples of HRNN(4) predictions for disaggregated indexes.*

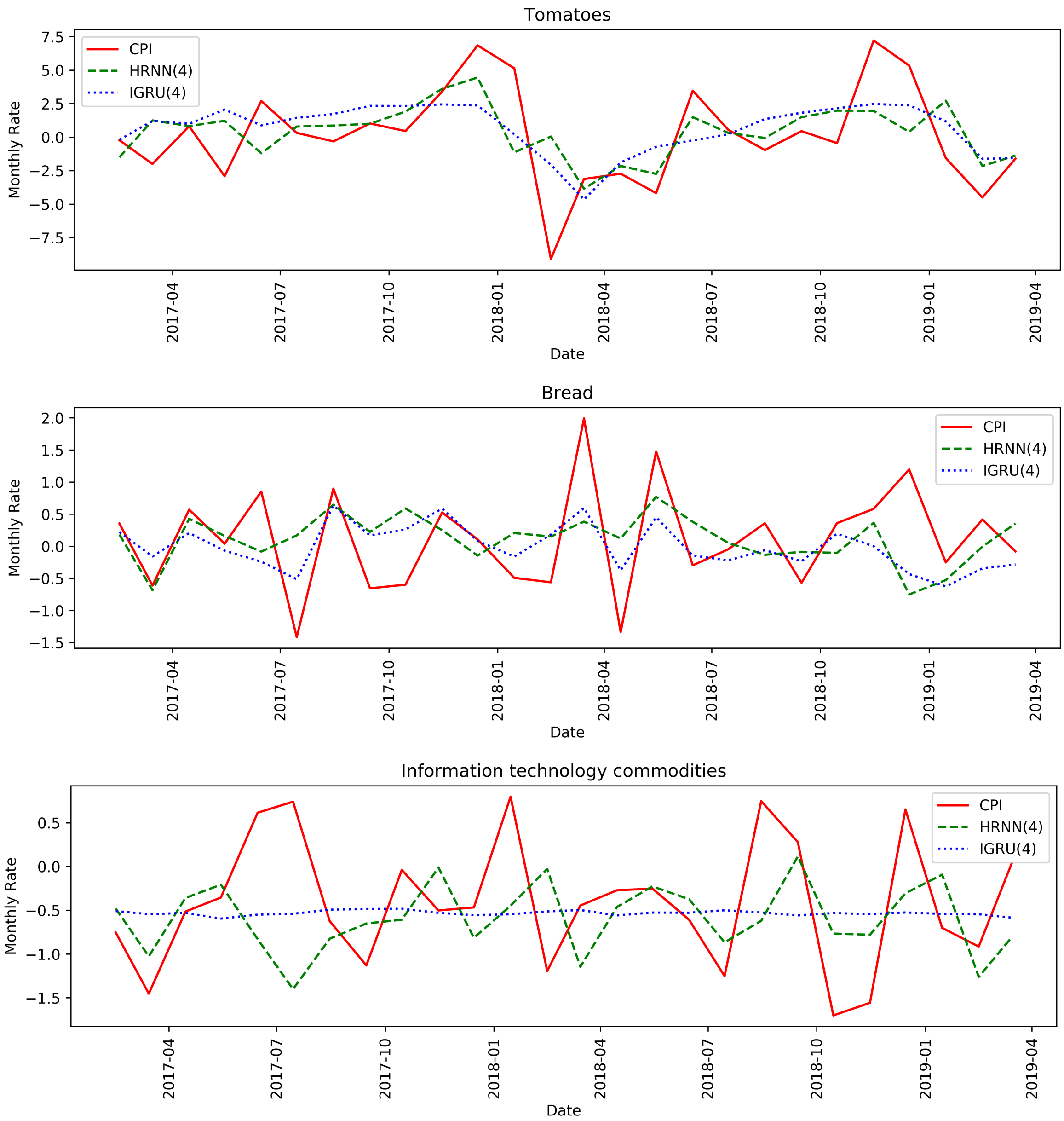

### 6.5. *HRNN Dynamics*

In what follows, we take a closer look at several characteristics of the HRNN model that result from the non-stationary nature of the CPI. The HRNN model is a deep learning hierarchical model that requires substantial training time depending on the available hardware. In this work, the HRNN model was trained once using the training dataset and evaluated on the test dataset as explained earlier. In order to investigate the potential benefit from retraining HRNN every quarter, we performed the following experiment: For a test-set period from 2001-2018, we retrained HRNN(4) after each quarter, each time adding the hierarchical CPI values of the last 3 months. Figure 8 presents the results of the this experiment. The dashed green line presents the RMSE of HRNN(4) with the "regular" training used in this work, while the dotted blue line presents the results of retraining HRNN every quarter. As expected, in most cases, retraining the model with additional data from the recent period improves the results. However, this improvement is moderate and the overall model quality is about the same.

**Figure 8.** *The Effect of Quarterly Retraining HRNN(4).*

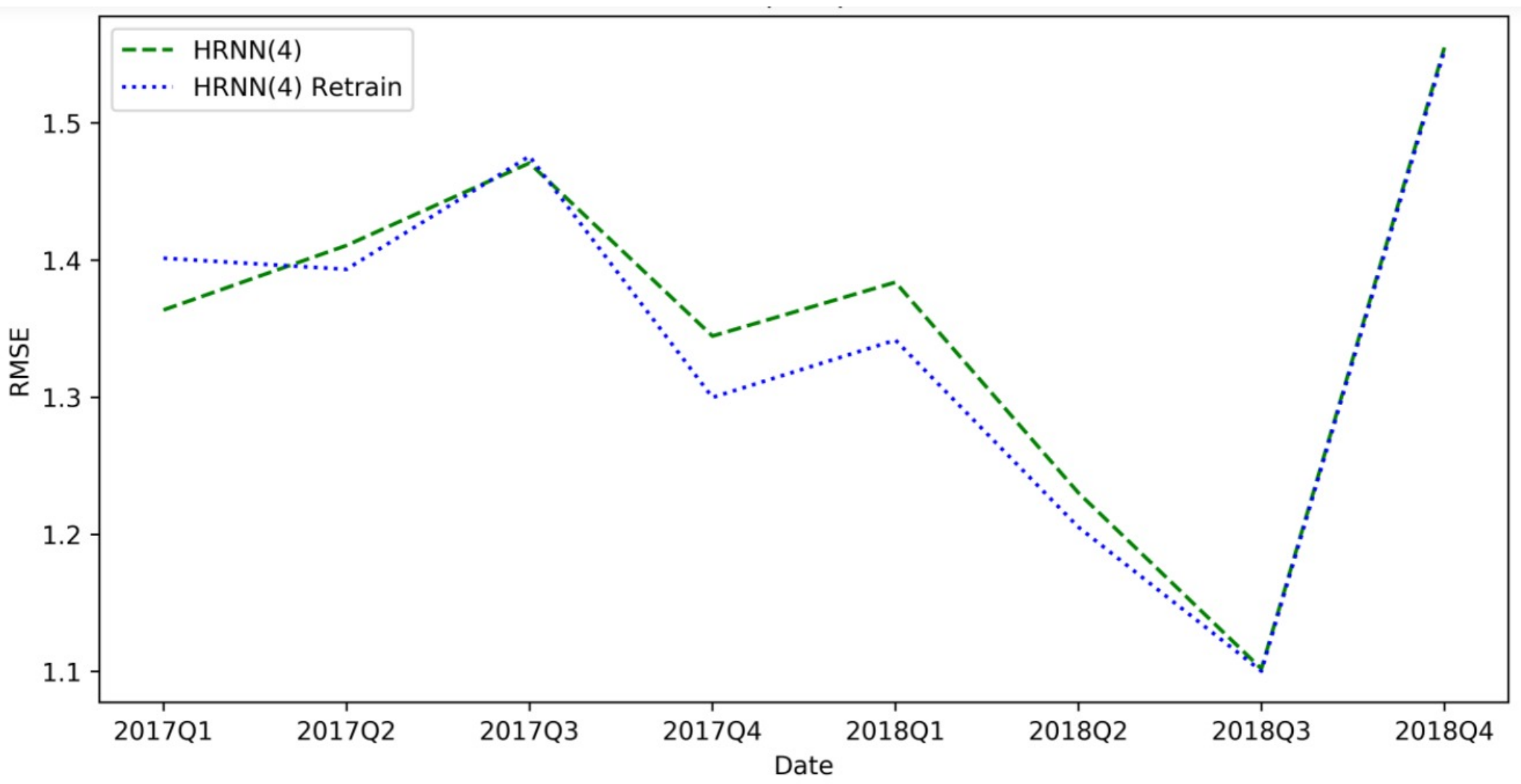

In order to study the GFC effect on HRNN's performance, we removed the data from 2008 onward and repeated the experiment of Table 2, using only the data from 1997 up to 2008. The results of this experiment are summarized in Table 6. In terms of RMSE, the gains of HRNN in Table 2 vary from 0.78 up to 0.8, in contrast to Table 6 where the gains vary from 0.83 to 0.93, revealing that during the turmoil of the GFC, when the demand for reliable and precise forecasting tools is enhanced, HRNN's forecasting abilities remain robust. In fact, its forecasting superiority was somewhat enhanced during the GFC when compared to the AR(1) baseline.

## 7. Concluding Remarks

Policymakers have a wide range of predictive tools at their disposal to forecast headline inflation: survey data, expert forecasts, inflation swaps, economic and econometric models, etc. However, policy institutions lack models and data to assist with

Table 6: Average Results on Disaggregated CPI Components Prior to The GFC

| Model Name* | RMSE per horizon AR(1)=1.00 | | | | | | Correlation (at horizon=0) | |
|---|---|---|---|---|---|---|---|---|
| | 0 | 1 | 2 | 3 | 4 | 8 | Pearson | Distance |
| AR(1) | 1.00 | 1.00 | 1.00 | 1.00 | 1.00 | 1.00 | 0.07 | 0.05 |
| AR(2) | 1.00 | 1.00 | 1.00 | 1.00 | 1.00 | 1.00 | 0.08 | 0.06 |
| AR(3) | 1.00 | 1.00 | 1.00 | 1.00 | 1.00 | 1.00 | 0.09 | 0.07 |
| AR(4) | 1.00 | 1.00 | 1.00 | 1.00 | 1.00 | 1.00 | 0.09 | 0.07 |
| AR-GAP(3) | 1.00 | 1.00 | 1.00 | 1.00 | 1.00 | 1.00 | 0.09 | 0.07 |
| AR-GAP(4) | 1.00 | 1.00 | 1.00 | 1.00 | 1.00 | 1.00 | 0.10 | 0.07 |
| RW(4) | 1.00 | 1.00 | 1.00 | 1.00 | 1.00 | 1.00 | -0.03 | -0.03 |
| Phillips(4) | 1.00 | 1.00 | 0.99 | 0.99 | 1.00 | 1.00 | -0.02 | -0.02 |
| VAR(1) | 1.04 | 1.04 | 1.04 | 1.05 | 1.05 | 1.06 | 0.04 | 0.03 |
| VAR(2) | 1.03 | 1.04 | 1.04 | 1.04 | 1.05 | 1.05 | 0.05 | 0.03 |
| VAR(3) | 1.03 | 1.03 | 1.03 | 1.04 | 1.04 | 1.05 | 0.06 | 0.03 |
| VAR(4) | 1.02 | 1.03 | 1.03 | 1.03 | 1.03 | 1.04 | 0.06 | 0.04 |
| LSTAR($\rho = 4, c = 2, \gamma = 0.3$) | 1.05 | 1.06 | 1.05 | 1.08 | 1.09 | 1.10 | 0.08 | 0.06 |
| RF(4) | 0.92 | 0.91 | 0.91 | 0.92 | 0.92 | 0.95 | 0.2 | 0.29 |
| GBT(4) | 0.91 | 0.92 | 0.91 | 0.93 | 0.92 | 0.97 | 0.18 | 0.34 |
| FC(4) | 0.99 | 0.99 | 1.00 | 1.00 | 1.02 | 1.05 | 0.11 | 0.08 |
| Deep-NN(4) | 0.94 | 0.95 | 0.94 | 0.94 | 0.94 | 0.95 | 0.15 | 0.32 |
| Deep-NN(4) + unemployment | 0.92 | 0.92 | 0.94 | 0.95 | 0.93 | 0.95 | 0.2 | 0.35 |
| S-GRU(4) | 1.05 | 1.09 | 1.09 | 1.10 | 1.09 | 1.10 | 0.09 | 0.07 |
| I-GRU(4) | 0.86 | 0.90 | 0.90 | 0.92 | 0.93 | 0.94 | 0.33 | 0.35 |
| KNN-GRU(1) | 0.94 | 0.96 | 0.96 | 0.96 | 0.97 | 0.98 | 0.10 | 0.07 |
| KNN-GRU(2) | 0.94 | 0.96 | 0.95 | 0.96 | 0.97 | 0.98 | 0.11 | 0.08 |
| KNN-GRU(3) | 0.93 | 0.96 | 0.95 | 0.96 | 0.96 | 0.98 | 0.11 | 0.08 |
| KNN-GRU(4) | 0.93 | 0.96 | 0.96 | 0.95 | 0.96 | 0.97 | 0.12 | 0.09 |
| HRNN(1) | 0.85 | 0.89 | 0.90 | 0.92 | 0.91 | 0.94 | 0.23 | 0.27 |
| HRNN(2) | 0.84 | 0.89 | 0.90 | 0.92 | 0.91 | 0.94 | 0.24 | 0.25 |
| HRNN(3) | 0.84 | 0.89 | 0.89 | 0.92 | 0.91 | 0.93 | 0.28 | 0.34 |
| **HRNN(4)** | **0.83** | **0.88** | **0.88** | **0.91** | **0.90** | **0.93** | **0.35** | **0.37** |

*Notes:* Average results across all 424 inflation indexes that make up the headline CPI. In contrast to Table 2, here we focus on results up to the GFC of 2008. The RMSE results are relative to the $AR(1)$ model and normalized according to its results, i.e., $\frac{RMSE_{Model}}{RMSE_{AR(1)}}$. Results are statistically significant according to Diebold-Mariano test with $p < 0.05$.

CPI components' forecasting, which are essential for a deeper understanding of the underlying dynamics. The understanding of disaggregated inflation trends can provide insight into the nature of future inflation pressures, their transitory factors (seasonal factors, energy, etc.), and other factors that influence market-makers and the conduct of monetary policy, among other decision-makers. Hence, our hierarchical approach uses endogenous historical data to forecast CPI at the disaggregated level, rather than forecasting headline inflation, even if it performs well (Ibarra, 2012).

The business cycle plays an important role in inflation dynamics, particularly through specific product classes. CPI inflation dynamics are sometimes driven by components unrelated to central bank policy objectives, such as food and energy prices, for example. A disaggregated CPI forecast provides a more accurate picture of the sources and features of future inflation pressures in the economy, which in turn improves policymakers' response efficiency. Indeed, forecasting sectoral inflation may improve the optimization problem faced by the central bank (Ida, 2020).

While similar headline inflation forecasts may correspond to various underlying economic factors, a disaggregated perspective allows understanding and analyzing the decomposition of these inflation forecasts at the sectoral or component level. Instead of disaggregating inflation to forecast the headline inflation (Stock and Watson, 2020), our approach allows policy and market makers to forecast specific sector and component prices, where information is less available: almost no component or sectoral-specific survey forecasts, expert forecast, or market-based forecasts exist. For instance, a central bank could use such modeling features to consider components that contribute to inflation (military, food, cigarettes, and energy) unrelated to its primary inflation objectives to improve their final assessment of their inflation forecasts. Sector-specific inflation forecasts should also inform economic policy recommendations at the sectoral level, and market makers can better direct and tune their investment strategies (Swinkels, 2018).

In traditional approaches for inflation forecasting, a theoretical or a linear model is often used, which inevitably biases the estimated forecasts. Our novel approach may overcome the usual shortcomings of traditional forecasts, giving policymakers new insights from a "different angle". Disaggregated forecasts include explanatory variables with hierarchies that reduce measurement errors at the component level. Additionally, our model structure attenuates component-specific residuals derived from each level and sector, resulting in improved forecasting. For all these reasons, we believe that HRNN can become a valuable tool for asset managers, policy institutions, and market makers lacking component-specific price forecasts critical to their decision processes.

The HRNN model was designed for predicting disaggregated CPI components, however we believe its merits may come useful in the prediction of other hierarchical time series such as GDP. In future work, we plan to investigate the performance of the HRNN model on additional hierarchical time series. Moreover, in this paper we focused mainly on endogenous models that do not consider other economic variables. HRNN can naturally be extended to include different variables as side information by changing the input for the GRU components to be a multi-dimensional time series (instead of a 1-dimensional vector). In future work, we plan to experiment with additional side information that can potentially improve the prediction accuracy. In particular, we plan to experiment with online price data as in Aparicio and Bertolotto (2020). Finally, we also plan to try to replace the RNNs in the model with neural self-attention (Shaw et al., 2018). Hopefully, this should lead to improved accuracy and better explainability through the analysis of attention scores (Hsieh et al., 2021).

## Appendix A Additional Tables and Figures

Table 7: Indexes Level 0 And 1

| Level | Index | Parent |
|---|---|---|
| 0 | All items | - |
| 1 | All items less energy | All items |
| 1 | All items less food | All items |
| 1 | All items less food and energy | All items |
| 1 | All items less food and shelter | All items |
| 1 | All items less food, shelter, and energy | All items |
| 1 | All items less food, shelter, energy, and used cars and trucks | All items |
| 1 | All items less homeowners costs | All items |
| 1 | All items less medical care | All items |
| 1 | All items less shelter | All items |
| 1 | Apparel | All items |
| 1 | Apparel less footwear | All items |
| 1 | Commodities | All items |
| 1 | Commodities less food | All items |
| 1 | Durables | All items |
| 1 | Education and communication | All items |
| 1 | Energy | All items |
| 1 | Entertainment | All items |
| 1 | Food | All items |
| 1 | Food and beverages | All items |
| 1 | Fuels and utilities | All items |
| 1 | Household furnishings and operations | All items |
| 1 | Housing | All items |
| 1 | Medical care | All items |
| 1 | Nondurables | All items |
| 1 | Nondurables less food | All items |
| 1 | Nondurables less food and apparel | All items |
| 1 | Other goods and services | All items |
| 1 | Other services | All items |
| 1 | Recreation | All items |
| 1 | Services | All items |
| 1 | Services less medical care services | All items |
| 1 | Services less rent of shelter | All items |
| 1 | Transportation | All items |
| 1 | Utilities and public transportation | All items |

*Note*: Levels and Parents of Indexes might change through time

Table 8: Indexes Level 2

| Level | Index | Parent |
|---|---|---|
| 2 | All items less food and energy | All items less energy |
| 2 | Apparel commodities | Apparel |
| 2 | Apparel services | Apparel |
| 2 | Commodities less food | Commodities |
| 2 | Commodities less food and beverages | Commodities |
| 2 | Commodities less food and energy commodities | All items less food and energy |
| 2 | Commodities less food, energy, and used cars and trucks | Commodities |
| 2 | Communication | Education and communication |
| 2 | Domestically produced farm food | Food and beverages |
| 2 | Education | Education and communication |
| 2 | Energy commodities | Energy |
| 2 | Energy services | Energy |
| 2 | Entertainment commodities | Entertainment |
| 2 | Entertainment services | Entertainment |
| 2 | Food | Food and beverages |
| 2 | Food at home | Food |
| 2 | Food away from home | Food |
| 2 | Footwear | Apparel |
| 2 | Fuels and utilities | Housing |
| 2 | Homeowners costs | Housing |
| 2 | Household energy | Fuels and utilities |
| 2 | Household furnishings and operations | Housing |
| 2 | Infants' and toddlers' apparel | Apparel |
| 2 | Medical care commodities | Medical care |
| 2 | Medical care services | Medical care |
| 2 | Men's and boys' apparel | Apparel |
| 2 | Nondurables less food | Nondurables |
| 2 | Nondurables less food and apparel | Nondurables |
| 2 | Nondurables less food and beverages | Nondurables |
| 2 | Nondurables less food, beverages, and apparel | Nondurables |
| 2 | Other services | Services |
| 2 | Personal and educational expenses | Other goods and services |
| 2 | Personal care | Other goods and services |
| 2 | Pets, pet products and services | Recreation |
| 2 | Photography | Recreation |
| 2 | Private transportation | Transportation |
| 2 | Public transportation | Transportation |
| 2 | Rent of shelter | Services |
| 2 | Services less energy services | All items less food and energy |
| 2 | Services less medical care services | Services |
| 2 | Services less rent of shelter | Services |
| 2 | Shelter | Housing |
| 2 | Tobacco and smoking products | Other goods and services |
| 2 | Transportation services | Services |
| 2 | Video and audio | Recreation |
| 2 | Women's and girls' apparel | Apparel |

*Note*: Levels and Parents of Indexes have changed over the years.

**Figure 9.** *Additional Examples of HRNN(4) predictions for disaggregated indexes*

**(a)** *Admission to movies, theaters, and concerts*

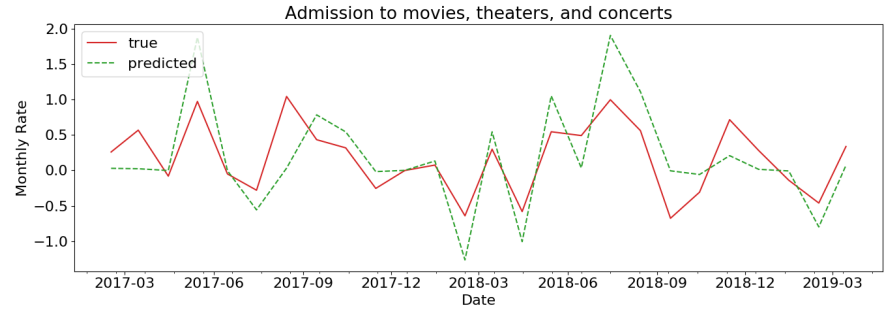

**(b)** *Alcoholic beverages*

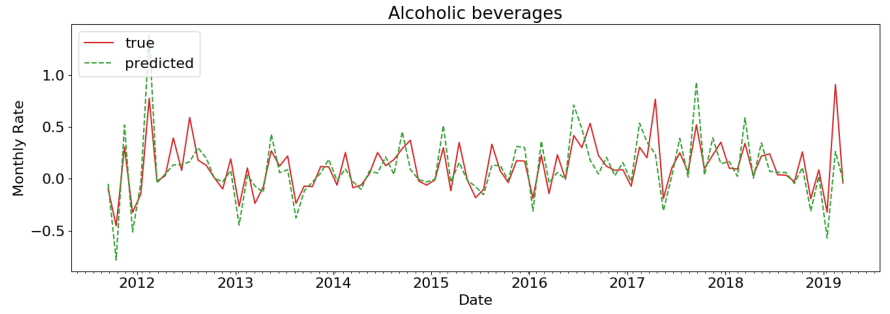

**(c)** *Bacon and related products*

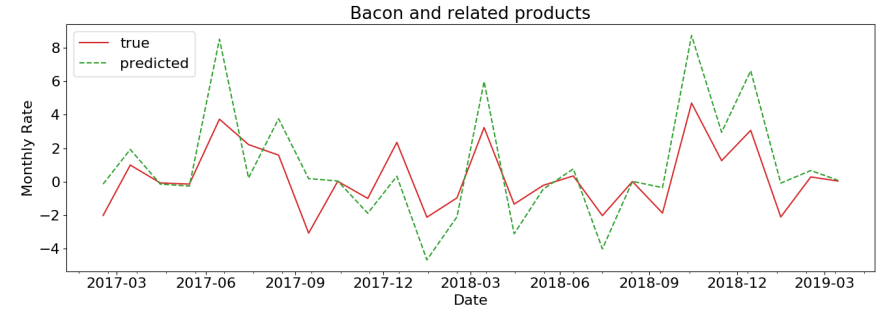

**(d)** *Education*

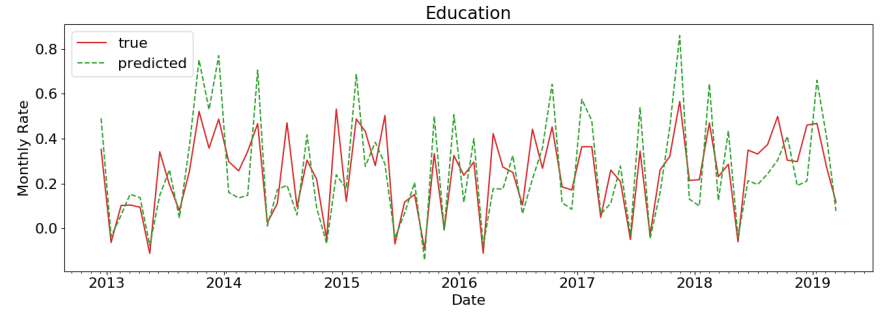

**(e)** *Film processing*

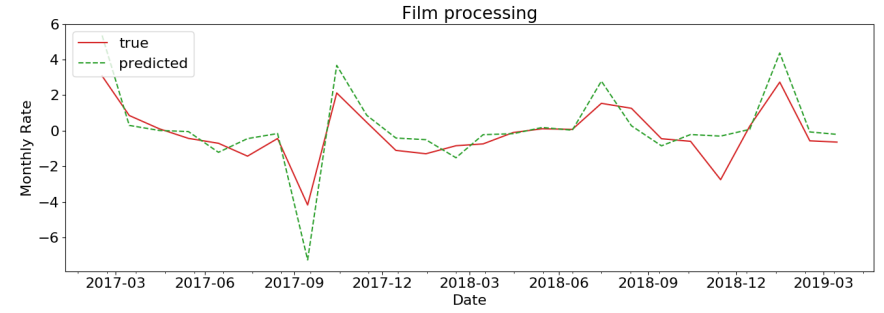

**(f)** *Financial services*

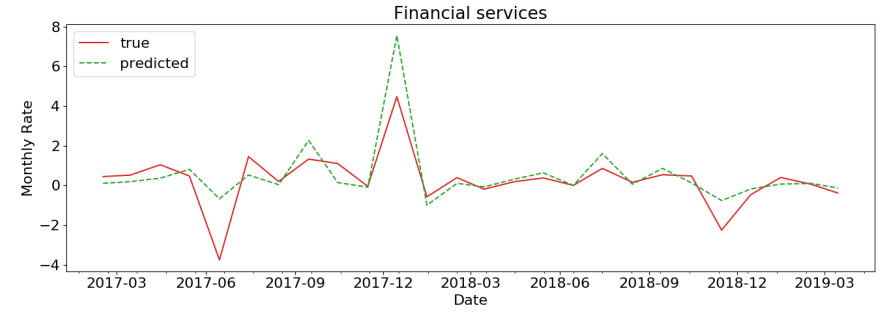

**(g)** *Fruits and vegetables*

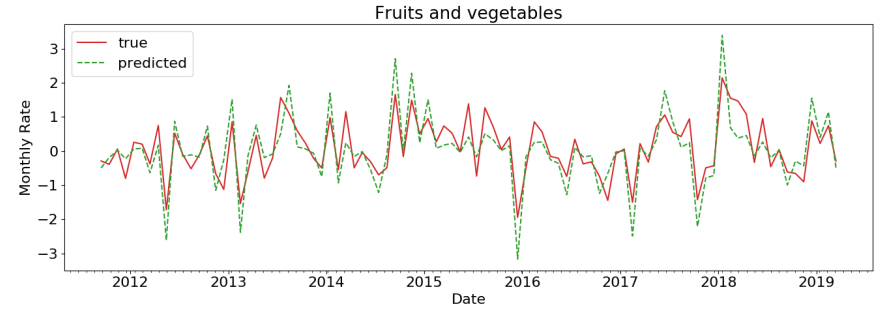

**(h)** *Gasoline, unleaded regular*

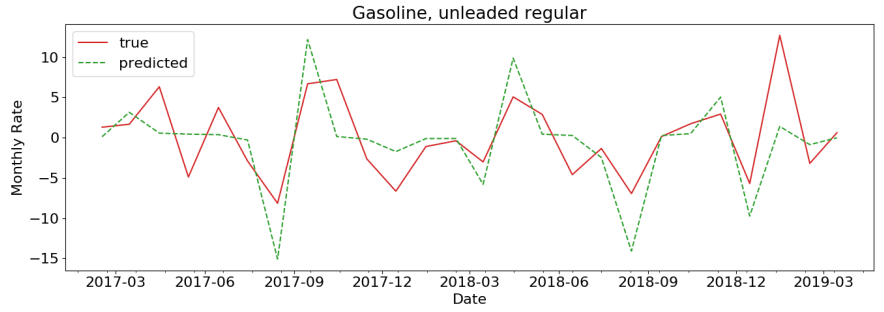

**(i)** *Haircuts and other personal care services*

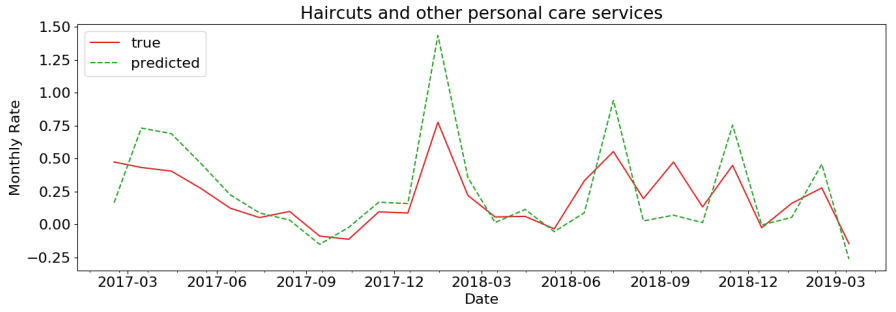

**(j)** *Hospital services*

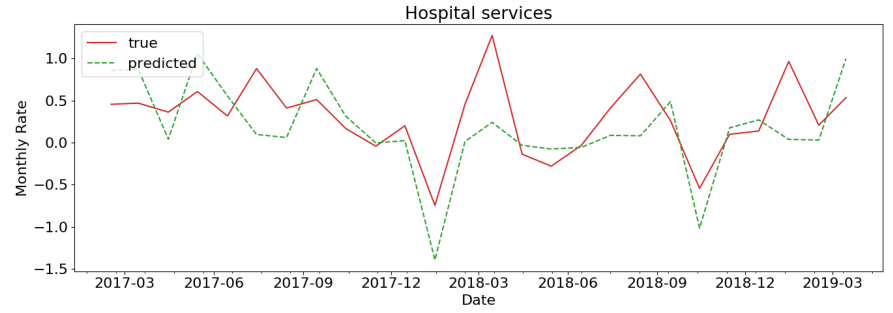

**(k)** *Household energy*

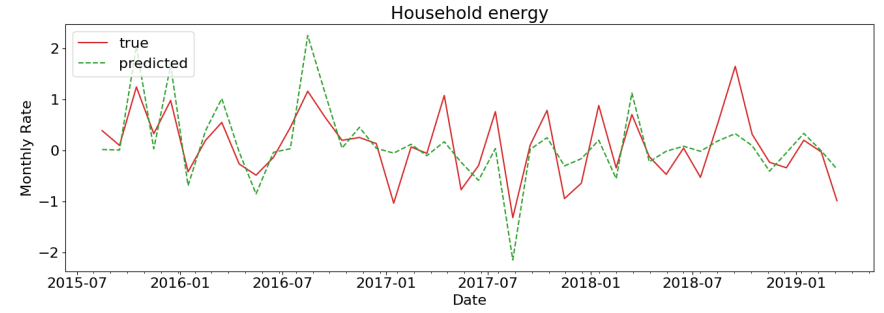

**(l)** *Household operations*

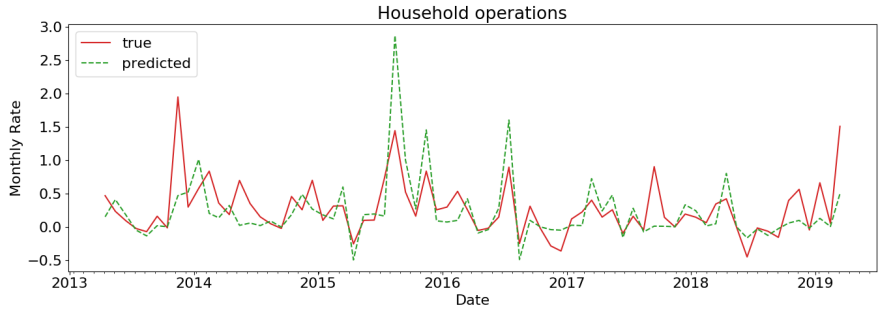

**Figure 10.** *Additional Examples of HRNN(4) predictions for disaggregated indexes (different hierarchies and sectors)*

**(a)** *Housing*

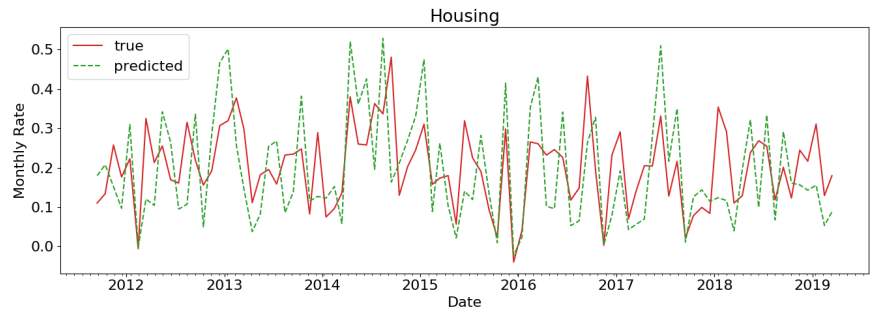

**(b)** *Intercity train fare*

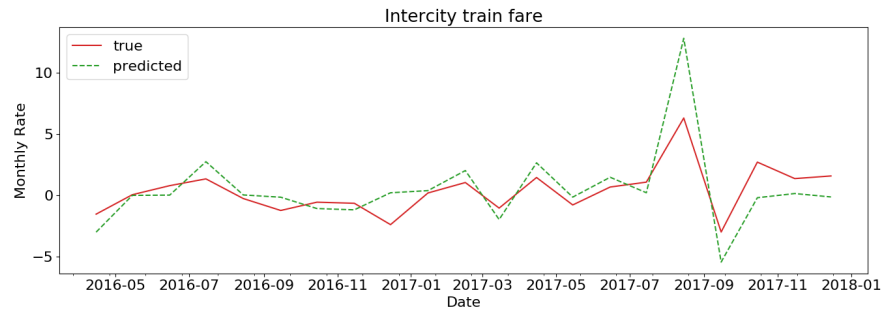

**(c)** *Jewelry*

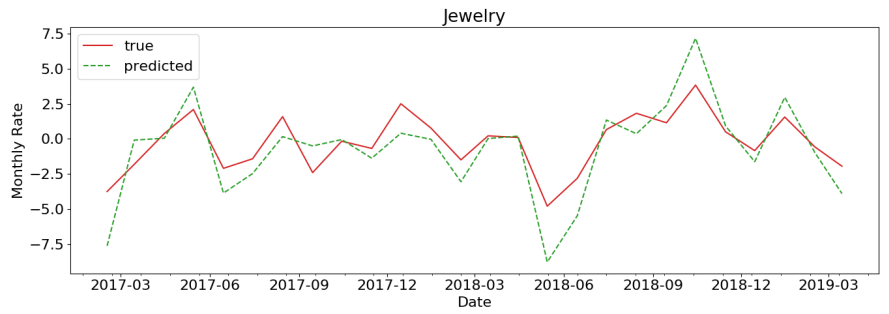

**(d)** *Medical care commodities*

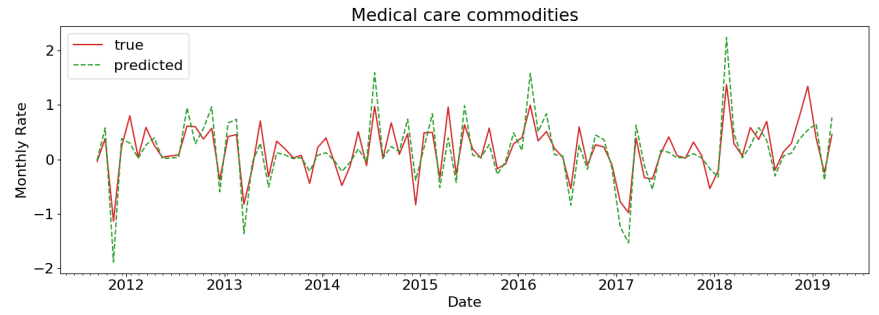

**(e)** *Medical care*

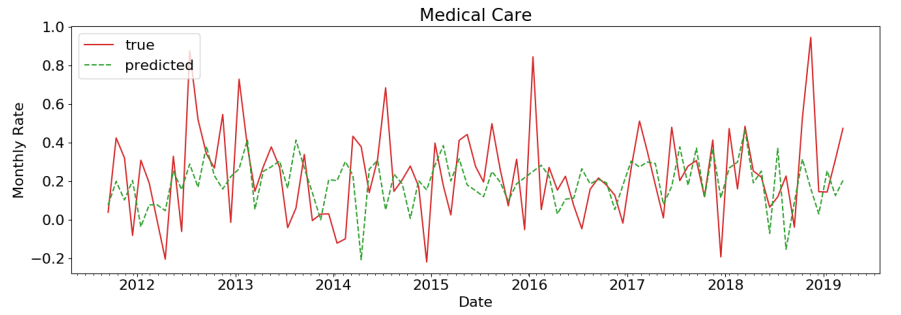

**(f)** *Motor oil, coolant, and fluids*

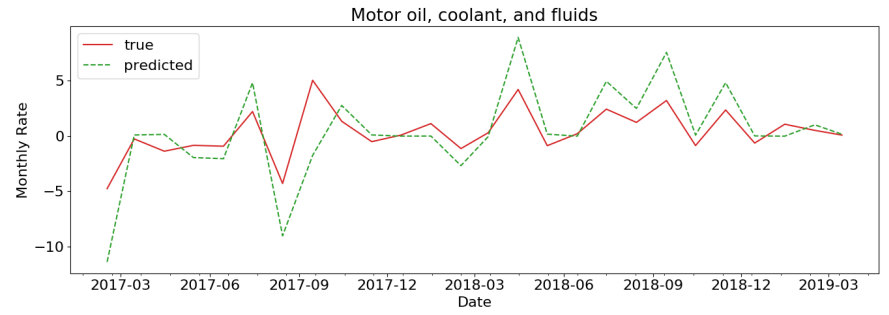

**(g)** *Motor vehicle insurance*

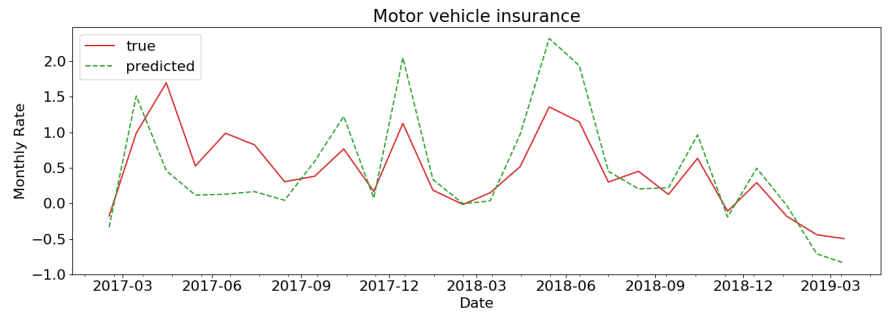

**(h)** *Nonprescription drugs*

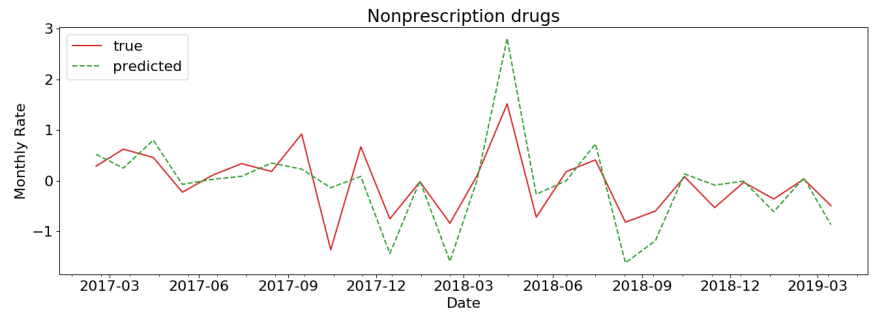

**(i)** *Private transportation*

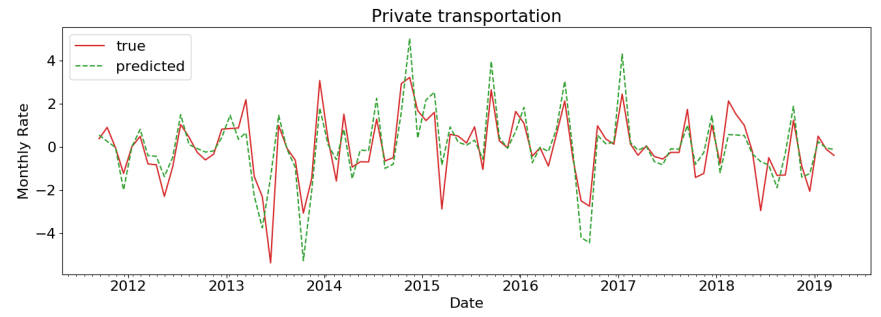

**(j)** *Sports equipment*

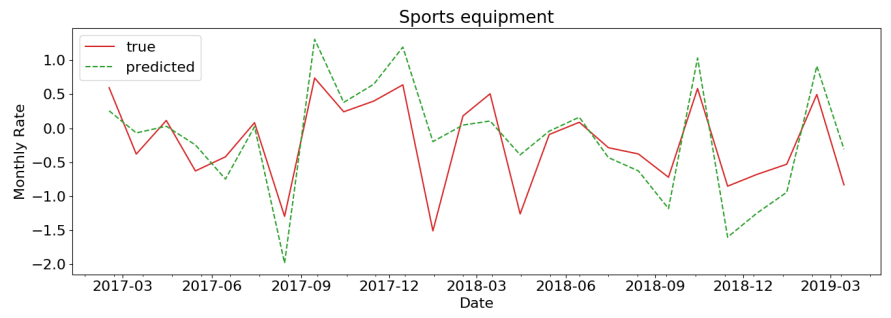

**(k)** *White bread*

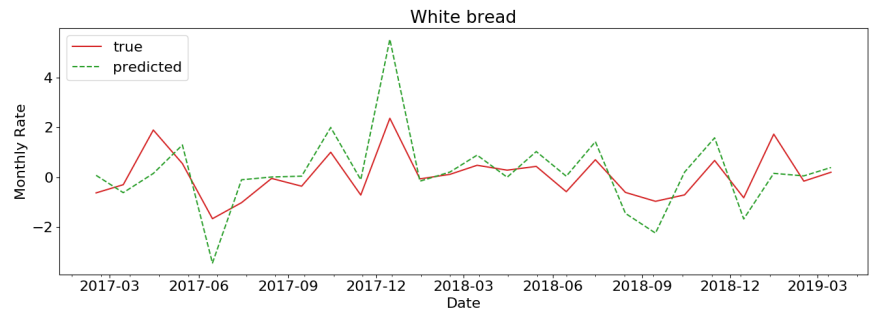

**(l)** *Women's apparel*

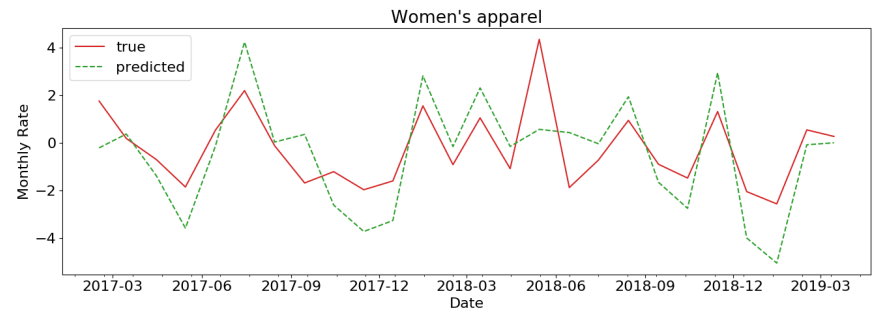

*Figures 13-24, indexes were selected from different hierarchies and sectors*